\documentclass[twocolumn, twocolappendix]{aastex701}

\usepackage{xspace}
\usepackage{amsmath}

\newcommand{\xmm}{XMM-\textit{Newton}\xspace}
\newcommand{\nustar}{\textit{NuSTAR}\xspace}
\newcommand{\suzaku}{\textit{Suzaku}\xspace}
\newcommand{\xrism}{\textit{XRISM}\xspace}
\newcommand{\resolve}{\textit{Resolve}\xspace}
\newcommand{\xtend}{\textit{Xtend}\xspace}
\shorttitle{Multi-epoch Detection of an Ultra-fast Inflow in ESP~39607}
\shortauthors{Peca et al.}
\graphicspath{{./}{./figures/}}

\defcitealias{peca25a}{P25}

\begin{document}

\title{Multi-epoch Detection of an Ultra-fast Inflow in ESP~39607: Evidence for an Accretion Cascade}

\author[orcid=0000-0003-2196-3298]{Alessandro Peca}
\affiliation{Eureka Scientific, 2452 Delmer Street, Suite 100, Oakland, CA 94602-3017, USA} 
\affiliation{Department of Physics, Yale University, P.O. Box 208120, New Haven, CT 06520, USA}
\affiliation{Universidad Diego Portales, Facultad de Ingenier\'ia y Ciencias, Instituto de Estudios Astrof\'isicos, Av. Ej\'ercito Libertador 441, Santiago, Chile}
\email[show]{peca.alessandro@gmail.com}

\author[orcid=0000-0002-7998-9581]{Michael J. Koss}
\affiliation{Eureka Scientific, 2452 Delmer Street, Suite 100, Oakland, CA 94602-3017, USA}
\affiliation{Space Science Institute, 4750 Walnut Street, Suite 205, Boulder, CO 80301, USA}
\email{mike.koss@eurekasci.com}

\author[0000-0003-1200-5071]{Roberto Serafinelli}
\affiliation{Universidad Diego Portales, Facultad de Ingenier\'ia y Ciencias, Instituto de Estudios Astrof\'isicos, Av. Ej\'ercito Libertador 441, Santiago, Chile}
\affiliation{INAF - Osservatorio Astronomico di Roma, Via Frascati 33, 00078, Monte Porzio Catone (Roma), Italy}
\email{roberto.serafinelli@mail.udp.cl}

\author[0000-0002-0745-9792]{C. Megan Urry}
\affiliation{Department of Physics, Yale University, P.O. Box 208120, New Haven, CT 06520, USA}
\affiliation{Yale Center for Astronomy \& Astrophysics, 52 Hillhouse Avenue, New Haven, CT 06511, USA}
\email{meg.urry@yale.edu}

\author[0000-0001-5231-2645]{Claudio Ricci}
\affiliation{Department of Astronomy, University of Geneva, ch. d’Ecogia 16, 1290, Versoix, Switzerland}
\email{Claudio.Ricci@unige.ch}

\author[0000-0001-5709-7606]{Keigo Fukumura}
\affiliation{Department of Physics and Astronomy, James Madison University, Harrisonburg, VA 22807, USA}
\email{fukumukx@jmu.edu}

\author[0000-0003-2754-9258]{Massimo Gaspari}
\affiliation{Department of Physics, Informatics and Mathematics, University of Modena and Reggio Emilia, 41125 Modena, Italy}
\email{massimo.gaspari@unimore.it}

\author[0000-0002-0273-218X]{Elias Kammoun}
\affiliation{Cahill Center for Astronomy \& Astrophysics, California Institute of Technology, 1216 East California Boulevard, Pasadena, CA 91125, USA}
\email{ekammoun@caltech.edu}

\author[0000-0003-3450-6483]{Alessia Tortosa}
\affiliation{INAF - Osservatorio di Astrofisica e Scienza dello Spazio di Bologna, via Piero Gobetti, 93/3, I-40129 Bologna, Italy}
\email{alessia.tortosa@inaf.it}

\author[0000-0002-5273-4634]{Giulia Cerini}
\affiliation{Jet Propulsion Laboratory, California Institute of Technology, 4800 Oak Grove Dr, Pasadena, CA 91109, USA}
\email{giulia.cerini@jpl.nasa.gov}

\author[0000-0001-9379-4716]{Peter G. Boorman}
\affiliation{Max-Planck-Institut für Extraterrestrische Physik, Gießenbachstraße 1, 85748 Garching, Germany}
\email{boorman@mpe.mpg.de}

\author[0000-0002-7962-5446]{Richard Mushotzky}
\affiliation{Department of Astronomy, University of Maryland, College Park, MD 20742, USA}
\affiliation{Joint Space-Science Institute, University of Maryland, College Park, MD 20742, USA}
\email{rmushotz@umd.edu}

\author[0000-0002-9948-0897]{Quirino D'Amato}
\affiliation{INAF - Istituto di Astrofisica e Planetologia Spaziali, Via del Fosso del Cavaliere, 00133 Roma, Italy}
\email{quirino.damato@inaf.it}


\begin{abstract}

We present simultaneous \xrism, \xmm, and \nustar observations of ESP~39607, a Seyfert~2 galaxy at $z = 0.201$. \xrism/\resolve reveals two absorption features near 4.7 and 4.9~keV in the observed frame, consistent with redshifted Fe\,\textsc{xxv} He$\alpha$/Fe\,\textsc{xxvi} Ly$\alpha$ absorption from gas inflowing at $v_\mathrm{in}\simeq0.16c$. The high velocity identifies the absorber as an ultra-fast inflow (UFI), which is detected at $\sim3$--$3.7\sigma$ across different methods and continuum models. Photoionization modeling yields $\log\xi/\mathrm{erg~s^{-1}~cm} \simeq 3.7$--$3.8$ and a column density $\log N_\mathrm{H,abs}/\mathrm{cm}^{-2} \simeq 23.2$--$23.8$, the latter depending on the assumed metallicity. The inflow velocity and line properties are consistent with those reported from two earlier \nustar epochs, indicating that similar inflowing material was present over a baseline of at least 2.2~yr in the source rest frame. Across all three epochs, the combined detection significance of the UFI is $5.3\sigma$. Given the dynamical timescale of a few days at the inferred radius, $R \simeq 49$--$77\,R_\mathrm{g}$, the multi-year evidence favors a scenario in which the inflow is continuously replenished, forming an accretion ``cascade'', rather than a single long-lived cloud. With an estimated mass inflow rate of $\dot{M}_{\mathrm{in}}\simeq0.3$--$1.4\,M_\odot$~yr$^{-1}$, depending on metallicity, and a ratio $\dot{M}_{\mathrm{in}}/\dot{M}_{\mathrm{acc}}\simeq0.2$--$1.0$, the inflow could supply a substantial fraction of the accretion needed to power the central active galactic nucleus.

\end{abstract}

\keywords{\uat{Accretion}{14} ---
\uat{Active galactic nuclei}{16} ---
\uat{Black hole physics}{159} ---
\uat{High-luminosity active galactic nuclei}{2034} ---
\uat{High resolution spectroscopy}{2096} ---
\uat{Photoionization}{2060} ---
\uat{Seyfert galaxies}{1447} ---
\uat{Supermassive black holes}{1663} ---
\uat{X-ray active galactic nuclei}{2035} ---
\uat{X-ray astronomy}{1810}}

\section{Introduction}

Active galactic nuclei (AGN) are among the most energetic phenomena in the Universe, powered by mass accretion onto supermassive black holes (SMBHs) at the centers of their host galaxies. The innermost regions of the accretion flow, within a few tens of gravitational radii ($R_\mathrm{g} = GM_\mathrm{BH}/c^2$) of the SMBH, are best probed through X-ray observations, which are sensitive to the hot, Comptonized corona and to the reprocessing of radiation by the accretion disc and circumnuclear material \citep[e.g.,][]{mushotzky93,urry95}. Over the past two decades, X-ray spectroscopy has revealed a rich phenomenology of ionized absorption features in AGN spectra, commonly interpreted as signatures of disc winds and outflows driven by e.g., radiation pressure, thermal gradients, or magnetic forces \citep[e.g.,][]{tombesi13,king15}.

Among the most extreme manifestations of AGN winds are ultra-fast outflows (UFOs), characterized by blueshifted Fe~K absorption features (primarily Fe\,\textsc{xxv} He$\alpha$ and Fe\,\textsc{xxvi} Ly$\alpha$ at rest-frame 6.70 and 6.97~keV, respectively) with velocities in the range $v_\mathrm{out} \sim 0.03$--$0.5c$ \citep[e.g.,][]{chartas02,reeves03,cappi06,tombesi10,braito18,serafinelli19,reeves20,braito22,matzeu23,gianolli24}. UFOs are detected in $\sim$30--40\% of local AGN \citep{tombesi10,gofford13}, indicating that they are not rare phenomena but a common phase during nuclear accretion. Because such winds can carry substantial mass, momentum, and kinetic power from sub-pc scales into the circumnuclear environment, they are considered a key channel through which accreting SMBHs may couple to their host galaxies, regulate gas cooling and star formation, and contribute to the broader AGN-host co-evolution framework \citep[e.g.,][]{king15,laha21,lanzuisi24}. Recent \xrism/\resolve observations are now extending this picture by resolving the velocity and ionization structure of high-velocity AGN winds in unprecedented detail \citep[e.g.,][]{xrism25_multiufo,xiang25,reeves26,xu25,noda25}.

In contrast, the opposite phenomenon, ultra-fast inflows (UFIs) identified through redshifted Fe~K absorption features, remains largely unexplored. Only a handful of UFI candidates have been reported in the literature to date \citep[e.g.,][]{dadina05,reeves05,longinotti07,giustini17,pounds18,pounds24,peca25a,dadina26}, with individual detections generally not exceeding $\sim$3$\sigma$ and with the same inflow rarely detected in more than one observation. This makes their confirmation particularly challenging. Yet UFIs are of particular interest because they may provide a direct view of non-standard accretion onto SMBHs. 
Several physical scenarios could produce such high-velocity inflows, all
of which depart from the standard picture of a smooth, steady, nearly
Keplerian accretion disc \citep{shakura1973}. For example, in the Chaotic Cold Accretion framework \citep[CCA;][]{gaspari13,gaspari17,gaspari20}, multiphase clouds condense out of the hot halo and rain stochastically toward the nucleus, as directly traced by redshifted molecular absorption \citep[e.g.,][]{tremblay16,rose19}. A persistent, continuously replenished inflow, or accretion ``cascade'', reaching sub-parsec scales would be a possible observational manifestation of this scenario. UFIs may alternatively arise in accretion flows guided by an ordered poloidal magnetic field \citep[e.g.,][]{fukumura07,fukumura26}. Other channels include disc tearing in misaligned or counter-rotating configurations \citep{king06,nixon12,dogan18}, as well as failed disc winds \citep[e.g.,][]{proga04} and aborted jets \citep[e.g.,][]{ghisellini04}, in which launched material fails to escape and falls back toward the SMBH. Discriminating between these scenarios requires high spectral resolution, broadband coverage, and multi-epoch observations.

An exceptional source for studying UFIs is ESP~39607, a highly luminous Seyfert~2 galaxy at $z = 0.201$ \citep{vettolani98,landi10,ricci17}. \citet[][hereafter \citetalias{peca25a}]{peca25a} reported the serendipitous discovery of a candidate UFI in the \nustar spectra of ESP~39607. An absorption feature consistently detected at $\sim$4.8~keV in two independent \nustar observations (obtained in May 2023 and August 2024) was interpreted as the blended contribution of Fe\,\textsc{xxv} He$\alpha$ and Fe\,\textsc{xxvi} Ly$\alpha$ absorption with an inflowing velocity of $v_\mathrm{in} \sim 0.15c$. With a combined detection significance above $4\sigma$, the inflow in ESP~39607 stands as one of the most compelling UFI candidates known and, to our knowledge, the only such candidate identified in a Type~2 AGN. It therefore offers a unique opportunity to investigate non-standard, extreme accretion dynamics in the immediate vicinity of its SMBH.
In this paper, we present new, simultaneous \xrism, \xmm, and \nustar\ observations of ESP~39607, providing the first high-resolution characterization of this candidate UFI together with broadband coverage of the AGN hosting it.

The paper is organized as follows. In Section~\ref{sec:data} we describe the observations and data reduction procedures. Section~\ref{sec:spectral} presents the spectral analysis, while in Section~\ref{sec:discussion} we discuss our results and their physical implications. We summarize our conclusions in Section~\ref{sec:conclusions}. Throughout this paper we adopt a flat $\Lambda$CDM cosmology with $H_0 = 70$~km\,s$^{-1}$\,Mpc$^{-1}$, $\Omega_\mathrm{m} = 0.3$, and $\Omega_\Lambda = 0.7$. Unless otherwise noted, uncertainties are quoted at the 90\% confidence level, and centroid energies of observed spectral features are given in the observed frame.

\section{Data} \label{sec:data}

\begin{figure}[ht]
    \centering
    \includegraphics[width=\columnwidth]{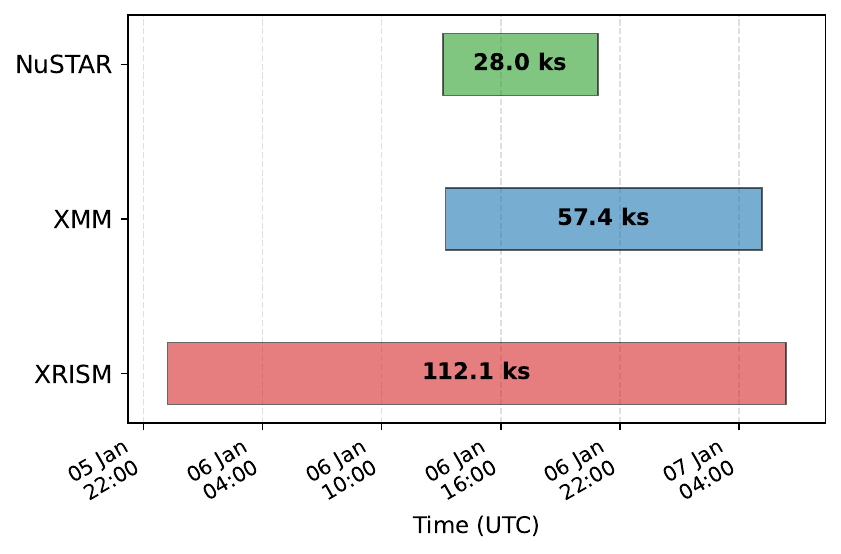}
    \caption{Timeline of the \xrism (red), \xmm (blue), and \nustar (green) observations. The \xrism observation started on 2026 January 5. The \xmm and \nustar observations began on 2026 January 6 and overlap with the \xrism exposure, providing simultaneous broadband coverage. The exposure times shown here are the raw, total exposure times.}
    \label{fig:timeline}
\end{figure}

ESP~39607 was observed simultaneously by \xrism \citep{xrism,xrism_new}, \xmm \citep{jansen01}, and \nustar \citep{harrison13} in January 2026. A timeline of the observations is shown in Figure~\ref{fig:timeline}.

\subsection{\nustar} \label{sec:nustar}

The \nustar observation (ObsID~91101649002) was performed on 2026 January 6. The data were reduced using the NuSTAR Data Analysis Software (\textsc{NuSTARDAS} v2.1.5), provided under HEASoft v6.35. We calibrated and cleaned the raw event files using the \texttt{nupipeline} routine. Source and background spectra were extracted for the two focal plane modules (FPMA and FPMB) using the \texttt{nuproducts} command. As in \citetalias{peca25a}, the source spectrum was extracted from a circular region of radius 60\arcsec, and the background from two circular regions of radius 75\arcsec\ on the same detector chip, avoiding chip gaps and nearby sources. Here and for the other imaging instruments, the extraction radii were chosen to optimize the signal-to-noise ratio \citep[e.g.,][]{peca25b}.
No solar flare activity or other quality flags were identified during this observation.

\subsection{\xmm} \label{sec:xmm}

The \xmm observation (ObsID~0960850101) was performed on 2026 January 6. Data were reduced using the \xmm Science Analysis System (SAS v22.1.0), following the standard SAS threads.
In this work, we focus on the EPIC-PN camera. We reduced the EPIC-PN data starting from the observation data files (ODFs), generating the calibration index and ODF summary files using \texttt{cifbuild} and \texttt{odfingest}, respectively, and then producing calibrated event lists with \texttt{epproc}. We filtered the event files for the flaring particle background following the standard SAS procedure. The response matrix (RMF) and ancillary response (ARF) files were generated using \texttt{rmfgen} and \texttt{arfgen}, respectively. Source spectra were extracted from a circular region of radius 22\arcsec\ centered on the source, and background spectra from a 75\arcsec\ circular region on the same CCD chip, away from CCD gaps and nearby sources.

\subsection{\xrism} \label{sec:xrism}

\xrism observed ESP~39607 starting on 2026 January 5 (ObsID~202035010). We used both \xrism instruments: the high-resolution micro-calorimeter spectrometer \resolve and the X-ray CCD imaging spectrometer \xtend. The data from both instruments were reduced following the \xrism\ Data Reduction ABC Guide v1.0.

\subsubsection{\resolve} \label{sec:resolve}

We reprocessed the raw data using the standard \xrism pipeline (\texttt{xapipeline}), adopting the most recent calibration database (CALDB 20250915) at the time of the analysis. Level-2 cleaned event files were produced following the standard pipeline. We applied the recommended screening for pulse rise time, event type, and status flags, and excluded events from pixel~27 to avoid calibration uncertainties. We applied a geomagnetic cut-off rigidity (COR) threshold of COR\,$>$\,6, which represents a reasonable compromise between exposure time and signal-to-noise ratio, and selected only the high-resolution (Hp) grade events. Spectra were extracted using the \textsc{xselect} task. We generated a ``large'' (L type) RMF file using the \texttt{rslmkrmf} command. An exposure map was produced with \texttt{xaexpmap}, and the ARF was generated using \texttt{xaarfgen}, assuming a point source at the source coordinates. Because the source is relatively faint for \resolve, the background contribution is non-negligible, especially below 3.5~keV and above 10~keV. We generated a high-statistics background spectrum following the HEASARC procedure for \resolve background simulations\footnote{\url{https://heasarc.gsfc.nasa.gov/docs/xrism/proposals/nxb_sky_rsl_bgd.html}}. Both the non-X-ray background (NXB) and the cosmic X-ray background (CXB) components were simulated according to these guidelines and combined into a single background spectrum. To make the per-channel Poisson fluctuations negligible and mimic a smooth background model, we simulated the background with an exposure of $10^{6}$~ks. The resulting spectrum was associated with the \resolve source spectrum as its background file and subsequently treated in the same manner as the background spectra of the other instruments.

\subsubsection{\xtend} \label{sec:xtend}

\xtend cleaned event files were processed with \textsc{xselect}, applying the recommended energy filter (PI\,$=$\,100--1665, or $\sim0.6$--10~keV), whose lower bound avoids contamination from cosmic-ray echo events, and the same COR\,$>$\,6 threshold used for \resolve. No solar flares or anomalous quality flags were identified. Source spectra were extracted from a circular region of radius 60\arcsec, and background spectra from two circular regions of radius 60\arcsec, all chosen to avoid CCD gaps. The RMF was generated with \texttt{xtdrmf}, an exposure map with \texttt{xaexpmap}, and the ARF with \texttt{xaarfgen}, assuming a point source at the source coordinates.

\subsection{Spectral binning} \label{sec:binning}

All spectra were grouped with the \texttt{ftgrouppha} tool, following the optimal binning scheme of \citet{kaastra16}. For the \resolve spectrum, given the relatively low source count rate, we additionally required a minimum of one count per bin to avoid empty channels. 

\section{Spectral fitting} \label{sec:spectral}

Spectral analysis was performed with XSPEC v12.15.0 \citep{xspec}, using the modified Cash statistic \citep[W-statistic;][]{cash79,wachter79} and fitting all datasets simultaneously. Since the source is X-ray obscured \citep[$\log N_\mathrm{H}/\mathrm{cm}^{-2} \simeq 23.5$,][]{ricci17,peca25a,peca25b}, we focus the analysis at energies above 2~keV, where the nuclear AGN emission dominates. The \xmm\ and \xtend\ data were fitted in the 2--10~keV band, while the \nustar\ data were fitted in the 3--35~keV band, above which the background dominates over the source emission. For \resolve, we restricted the fit to the 3.5--10~keV band, again excluding energies where the background dominates. The corresponding net exposures and count rates are 45.7~ks and $(4.78\pm0.10)\times10^{-2}$~counts~s$^{-1}$ for \xmm/PN, 27.9~ks and $(3.34\pm0.12)\times10^{-2}$~counts~s$^{-1}$ for \nustar/FPMA, 27.7~ks and $(3.17\pm0.12)\times10^{-2}$~counts~s$^{-1}$ for \nustar/FPMB, 94.3~ks and $(1.10\pm0.04)\times10^{-2}$~counts~s$^{-1}$ for \resolve, and 91.9~ks and $(1.96\pm0.05)\times10^{-2}$~counts~s$^{-1}$ for \xtend.

The analysis proceeds as follows. We first fit the broadband continuum using three independent models, finding clear absorption residuals in all cases (Section~\ref{sec:continuum}). We then perform a blind search for narrow absorption features and characterize the primary, most prominent feature phenomenologically (Section~\ref{sec:gaussians}). Next, we replace the phenomenological description with a more physical photoionization model to infer the properties of the absorber (Section~\ref{sec:xstar}). Finally, we examine the weaker secondary absorption features (Section~\ref{sec:secondary}) and the neutral Fe~K$\alpha$ emission profile (Section~\ref{sec:feka}). 

\subsection{Baseline continuum models} \label{sec:continuum}

We tested three physically motivated baseline models for the obscured AGN emission to assess the robustness of our results against different modeling assumptions. For each model, we also included an unobscured component representing leaked or scattered nuclear emission. Its spectral shape was tied to that of the primary continuum, while its relative normalization defined the scattering fraction. The individual model implementations are described below. Photon index and line-of-sight column density were linked across the model components.
In all models, we included a multiplicative constant to account for cross-calibration differences between instruments and a Galactic absorption component at the source position fixed at $N_\mathrm{H,Gal}=2.6\times10^{20}$~cm$^{-2}$ \citep{kalberla05}. We assumed a high-energy cut-off of $E_\mathrm{cut}=200$~keV, representative of the values observed in the local AGN population \citep[e.g.,][]{ricci17}. 

As our first model, we adopted the decoupled configuration of \textsc{MYTorus} \citep{mytorus,yaqoob12,yaqoob24}, which allows the column density of the line-of-sight transmitted continuum to differ from that associated with the reprocessed continuum and fluorescent-line components. In this configuration, the zeroth-order continuum is treated as a purely line-of-sight component, making its inclination angle a dummy parameter. We therefore fixed the inclination of the line-of-sight zeroth-order component to $90\degr$, following the standard prescription, so that its column density directly represents the line-of-sight obscuration. We initially included the scattered and line components at both $90\degr$ and $0\degr$, with their relative normalizations allowed to vary. After verifying that the $0\degr$ components converged to very low values consistent with zero, we fixed their normalizations to zero, while the $90\degr$ scattered and line components were fixed to unity because their best-fit values were consistent with one and largely unconstrained. The unobscured leaked/scattered emission was described with a secondary cut-off power law with $E_\mathrm{cut}=200$~keV, matching the intrinsic continuum. The column density associated with the reprocessed emission can therefore be interpreted as an effective global/average column ($N_\mathrm{H,glob}$), allowing the setup to approximate a non-uniform or patchy torus geometry. We found $N_\mathrm{H,glob}\simeq(6\pm3)\times10^{23}$~cm$^{-2}$, a factor of $\sim$1.8 higher than the line-of-sight value reported in Table~\ref{tab:baseline}, although the two quantities remain consistent within the uncertainties. This hints at a non-uniform obscurer \citep[e.g.,][]{yaqoob12,lamassa14}, with either our line of sight intercepting a lower-density region of the torus or the material producing the reflection being distinct from that responsible for the obscuration \citep{traina21,torres-alba23,pizzetti25}.

As a second model, we tested \textsc{X-skirtor}, an X-ray torus model calculated with the radiative-transfer code \textsc{SKIRT} \citep{skirtor} and developed for high-resolution \resolve\ spectroscopy. The model describes a uniform wedge torus of cold gas with a free covering factor, and includes bound-electron scattering and the intrinsic shapes of the fluorescent lines, sampled at the \resolve\ spectral resolution. We modeled the transmitted and reprocessed emission with their \texttt{dir} and \texttt{rpc} tables, respectively, with all parameters tied between the two components, to maintain self-consistency. The dedicated \resolve-resolution tables were applied to the \resolve\ spectrum and the corresponding CCD-resolution tables to the other instruments, with their parameters linked. We fixed the covering factor to its best-fit value of 0.45 (see below). The unobscured leaked/scattered emission was described with a secondary cut-off power law as in \textsc{MYTorus}.

Finally, for direct comparison with \citetalias{peca25a}, we considered the clumpy torus model \textsc{uxclumpy} \citep{buchner19}. The model represents the obscurer as a population of clouds with an angular dispersion and an optional inner Compton-thick ring, controlled by the \texttt{TORsigma} and \texttt{CTKcover} parameters, respectively, and a geometry calibrated to reproduce the obscuration and eclipse statistics observed in AGN. When left free, \texttt{CTKcover} was not required by the fit and was consistent with zero. We therefore fixed \texttt{CTKcover}=0. We fixed \texttt{TORsigma} to its best-fit value of $20.7$, corresponding to a covering factor of $\approx0.4$ according to the tabulated \textsc{uxclumpy} geometries of \citet{boorman24_hexp}, and consistent with the value adopted for \textsc{X-skirtor}. We modeled the transmitted and reprocessed emission with the \texttt{cutoff-transmit} and \texttt{cutoff-reflect} tables, respectively, with their parameters linked to maintain self-consistency, as done for \textsc{X-skirtor}. The unobscured leaked/scattered component was described with the dedicated \texttt{cutoff-omni} table, with its spectral parameters also linked to the primary continuum.

Geometrical parameters are notoriously difficult to constrain through X-ray spectral fitting of obscured AGN \citep[e.g.,][]{buchner21,saha22,kallova24,boorman24}. For \textsc{X-skirtor} and \textsc{uxclumpy}, the covering factor and \texttt{TORsigma}, respectively, are not individually well constrained, but converge to mutually consistent geometries with covering factor $\approx0.4$. Similarly, when left free, the inclination is consistent with the value of $i=70\degr$ obtained by \citet{peca25b} from an extensive broadband \nustar\ and \suzaku\ analysis of this source. Given these weak individual constraints, we fixed the \textsc{X-skirtor} covering factor and the \textsc{uxclumpy} \texttt{TORsigma} to the values quoted above, and the inclination of both models to $70\degr$, thereby limiting degeneracies with the other model parameters. At these adopted configurations, the relative normalization of the reprocessed and transmitted components is consistent with unity when allowed to vary, as expected for physically self-consistent torus models, further supporting the assumed geometries.
We also verified that the adopted geometries provide an adequate description when excluding the main absorption feature at $\sim4.8$~keV (Section~\ref{sec:gaussians}) from all spectra and the neutral Fe~K$\alpha$ region from \resolve, and that varying the fixed geometric parameters over plausible values does not significantly affect the detection of the $\sim4.8$~keV feature.

\begin{figure*}[ht]
\centering
\includegraphics[width=0.49\textwidth]{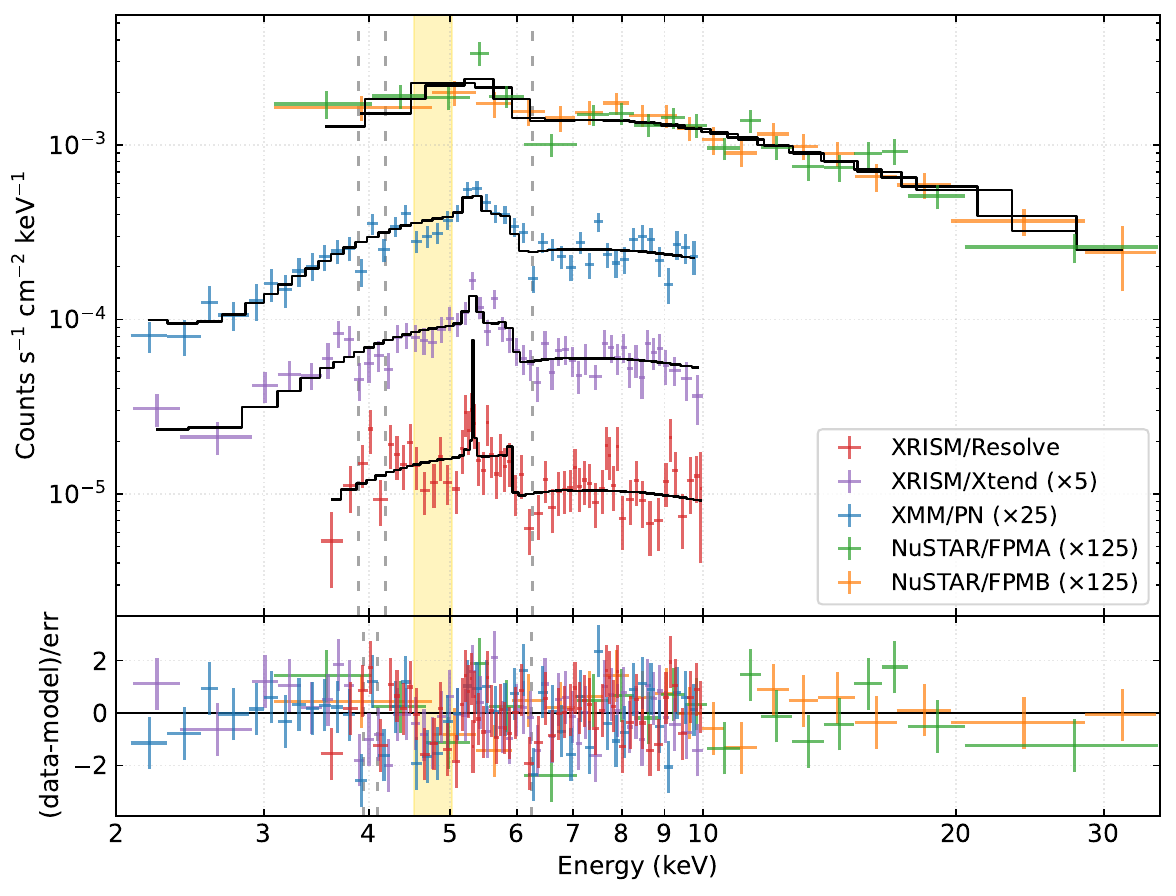}
\hfill
\includegraphics[width=0.49\textwidth]{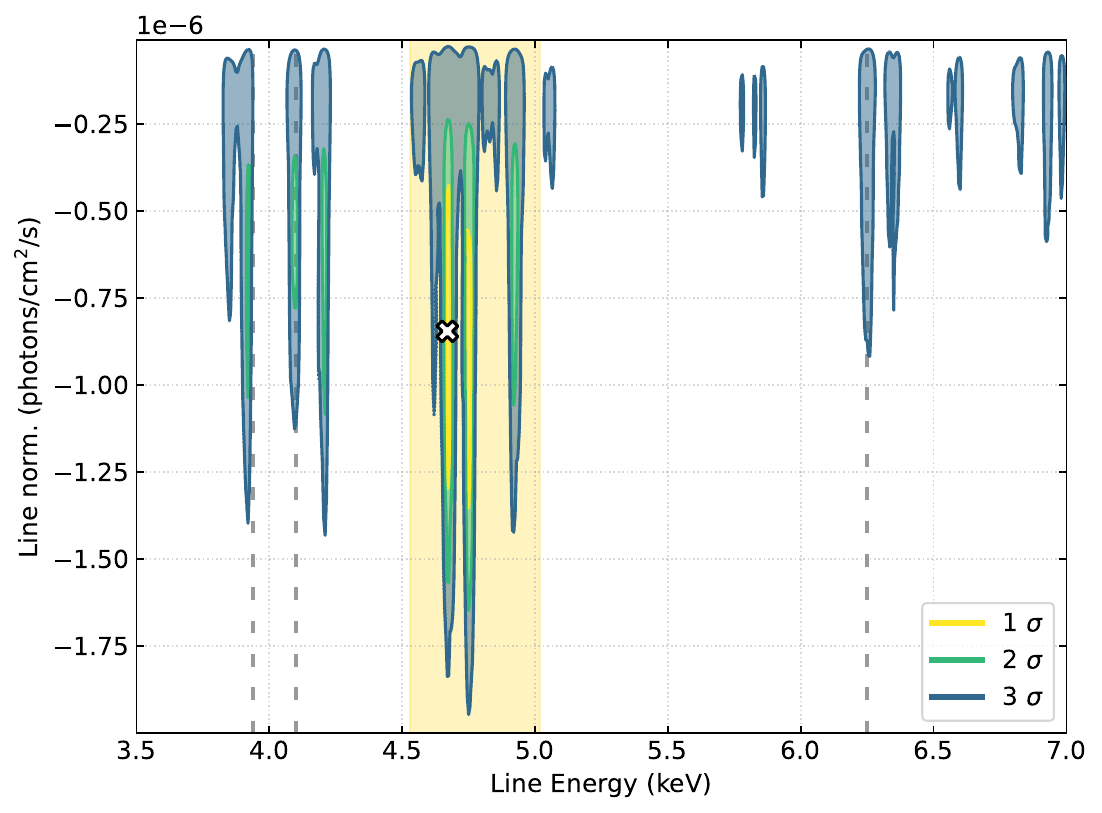}
\caption{\textit{Left:} Simultaneous \xrism/\resolve (red), \xrism/\xtend (purple), \xmm/PN (blue), \nustar/FPMA and FPMB (green and orange) spectra of ESP~39607 fitted with the baseline \textsc{MYTorus} continuum model. The bottom panel shows the residuals. The normalization of all but \resolve spectra are shifted as indicated in the legend for visual clarity.
\textit{Right:} Results from the blind absorption scan over line normalization and energy.  The white cross marks the best-fit value. 1, 2, and $3\sigma$ confidence contours are shown in yellow, green, and blue, respectively. In both panels, the gold shaded band marks the energy interval containing the primary UFI pair, while the gray dashed lines indicate the energies of the secondary UFI and UFO candidates. All energies are shown in the observed frame.}
\label{fig:noabs}
\end{figure*}

The best-fit results are summarized in Table~\ref{tab:baseline}, while the left panel of Figure~\ref{fig:noabs} shows the \textsc{MYTorus} fit. The three models returned consistent results, with observed 2--10~keV fluxes $F_{2\text{--}10}^\mathrm{obs}\simeq9.3\times10^{-13}$~erg~s$^{-1}$~cm$^{-2}$, intrinsic 2--10~keV luminosities $\log L_{2\text{--}10}^\mathrm{intr}/\mathrm{erg~s^{-1}}\simeq44.45$--44.53, and line-of-sight obscuring columns $N_\mathrm{H}\simeq(34.2$--$36.0)\times10^{22}$~cm$^{-2}$. The photon index spans a somewhat wider range, from $\Gamma\simeq1.52$ to $1.69$, although the three values remain consistent within their uncertainties. The apparently higher scattering fraction returned by \textsc{uxclumpy} arises from the different prescription used to model the scattered emission. Replacing the \texttt{omni} table with the scattered cut-off power law adopted for the other two models reduces the \textsc{uxclumpy} scattering fraction to $\sim2\%$, consistent with the other models, while yielding a statistically equivalent fit and leaving the continuum parameters unchanged. We retain the \texttt{omni} table as the self-consistent \textsc{uxclumpy} prescription for the scattered emission. In all cases, however, absorption residuals remain evident in the data (Figure~\ref{fig:noabs}), motivating the absorption modeling described in Sections~\ref{sec:gaussians} and~\ref{sec:xstar}.

\begin{deluxetable*}{lccccccc}
\tabletypesize{\small}
\tablecaption{Best-fit parameters for the three baseline continuum models. From left to right: Model; Photon index; line-of-sight column density in units of $10^{22}$~cm$^{-2}$; scattering fraction; observed 2--10~keV flux in units of $10^{-13}$~erg~s$^{-1}$~cm$^{-2}$; $\log$ of the intrinsic, rest-frame 2--10~keV luminosity; C-statistic over degrees of freedom; and AIC value. The AIC is defined as $\mathrm{AIC}=-2\ln\mathcal{L}+2k$ \citep{akaike74}, where $\mathcal{L}$ is the maximum likelihood and $k$ is the number of free parameters. \label{tab:baseline}}
\tablewidth{0pt}
\tablehead{
\colhead{Model} & \colhead{$\Gamma$} & \colhead{$N_\mathrm{H}$} & \colhead{$f_\mathrm{scat}$ (\%)} & \colhead{$F_{2\text{--}10}^\mathrm{obs}$} & \colhead{$\log L_{2\text{--}10}^\mathrm{intr}$} & \colhead{$\mathcal{C}$/dof} & \colhead{AIC}
}
\startdata
\textsc{MYTorus}   & $1.54^{+0.12}_{-0.13}$ & $34.2^{+3.6}_{-3.4}$ & $2.9^{+0.8}_{-0.6}$ & $9.3^{+0.3}_{-0.9}$ & $44.46^{+0.16}_{-0.19}$ & 1020.1/1249 & 1036.1 \\
\textsc{X-skirtor}  & $1.52^{+0.08}_{-0.12}$ & $35.5^{+2.7}_{-3.2}$ & $2.7^{+0.8}_{-0.6}$ & $9.3^{+0.4}_{-0.8}$ & $44.45^{+0.17}_{-0.15}$ & 1019.6/1250 & 1033.6 \\
\textsc{uxclumpy} & $1.69^{+0.09}_{-0.11}$ & $36.0^{+3.2}_{-3.8}$ & $4.7^{+1.4}_{-1.2}$ & $9.3^{+0.3}_{-1.0}$ & $44.53^{+0.11}_{-0.14}$ & 1031.6/1250 & 1045.6 \\
\enddata
\end{deluxetable*}

The relative cross-calibration constants were determined by fixing the \nustar/FPMA and FPMB normalizations to unity and leaving the PN, \resolve, and \xtend\ constants free to vary. Across the three baseline continuum models, we found $C_\mathrm{PN}=0.87\pm0.06$, $C_\mathrm{Resolve}=0.91\pm0.07$, and $C_\mathrm{Xtend}=1.03\pm0.07$, broadly consistent with the cross-calibration offsets reported by \citet{xrism_xcal}. These constants remain stable across the models, including those with residual absorption components, varying by at most $\sim1.5\%$.

\subsection{Phenomenological modeling} \label{sec:gaussians}

To characterize the absorption residuals present in all baseline continuum models, we added a Gaussian absorption line to each model and performed two-dimensional XSPEC \texttt{steppar} scans over line centroid energy and normalization. For these scans, we fixed the intrinsic line width at $\sigma=10$~eV and verified that this choice does not affect the results \citep[see also][]{reeves26b}. The centroid energy and normalization were stepped across the 3.5--10~keV band, with the energy sampled in steps of 5~eV, comparable to the spectral resolution of \resolve. The resulting confidence contours (Figure~\ref{fig:noabs}, right panel) reveal a complex pattern of absorption-like residuals, with the most significant structures concentrated in three observed-frame bands: 3.8--4.2~keV, 4.5--5.0~keV, and around 6.3~keV.
To interpret these residuals, we adopted the line pattern expected for highly ionized Fe K absorption, with iron favored by its high cosmic abundance and the prominence of Fe K transitions in AGN X-ray spectra \citep[e.g.,][]{reeves05,longinotti07}. The strongest Fe\,\textsc{xxv} He$\alpha$ and Fe\,\textsc{xxvi} Ly$\alpha$ transitions provide a characteristic energy separation. Fe\,\textsc{xxv} He$\alpha$ has a rest-frame energy of 6.700~keV (the resonance transition of the He-like triplet), while Fe\,\textsc{xxvi} Ly$\alpha$ has an effective rest-frame energy of 6.966~keV (the weighted average of the doublet at 6.952 and 6.973~keV), giving a rest-frame separation of $\Delta E_\mathrm{rest}=0.266$~keV.

Within this framework, in the two lower-energy bands we identify two pairs of features, each with a separation consistent with the redshifted Fe\,\textsc{xxv} He$\alpha$/Fe\,\textsc{xxvi} Ly$\alpha$ spacing: a stronger pair at observed-frame $\sim$4.7 and 4.9~keV ($\sim$5.6 and 5.9~keV in the source rest frame), which we refer to as the primary pair, and a weaker pair at $\sim$3.9 and 4.1~keV ($\sim$4.7 and 4.9~keV in the source rest frame), which we refer to as the secondary pair. Both pairs are recovered consistently across all three baseline continuum models. The third band, in contrast, shows a single feature at $\sim$6.26~keV ($\sim$7.5~keV in the source rest frame), with a possible companion not clearly identified in all models. We therefore associate this region with a tentative single absorption line. The identification of the two pairs as Fe\,\textsc{xxv} He$\alpha$ and Fe\,\textsc{xxvi} Ly$\alpha$ absorption from redshifted ionized gas is further supported by the physically motivated \textsc{xstar} absorber grids described in Section~\ref{sec:xstar}. Additionally, the primary pair matches the line-identification pattern of the UFI candidate previously reported by \citetalias{peca25a}, and its centroid energies are consistent within the uncertainties with the blended-feature centroids measured in the earlier epochs (Appendix~\ref{app:linechar_var}). In what follows, we focus on the primary pair, while the weaker secondary pair and the tentative single line are discussed in Section~\ref{sec:secondary}.

We tested several phenomenological models for the primary pair against each of the three baseline continuum models. The full set of model comparisons is reported in Appendix~\ref{app:fullgrid}, while the principal results are summarized here and in Table~\ref{tab:modelcomp}. 
We first modeled the feature with two narrow Gaussian absorption lines, fixing their widths at $\sigma=10$~eV, comparable to the effective \resolve\ resolution for the adopted spectral binning. When allowed to vary, the widths remain poorly constrained and do not significantly improve the fit. Allowing both centroid energies to vary freely yields $\Delta\mathcal{C}/\Delta\mathrm{dof}\simeq18$--22/4, depending on the continuum model, with best-fit energies of $\simeq4.67$ and $\simeq4.92$~keV (see Appendix~\ref{app:linechar_var}).
Because the \resolve\ spectrum is relatively faint and provides the only data capable of resolving the two features, we consider the physically motivated linked-pair solution to be more robust. We implement this constraint by assigning both lines a common absorber redshift, so their rest-frame centroid separation of $\Delta E_\mathrm{rest}=0.266$~keV is preserved consistently into the observed frame. This linked-pair model gives $\Delta\mathcal{C}/\Delta\mathrm{dof}\simeq15$--18/3 (Table~\ref{tab:modelcomp}). Under the Gaussian approximation \citep[e.g.,][]{tozzi06}, the fit improvements correspond to $3.1$--$3.5\sigma$. The same model comparisons yield $\Delta\mathrm{AIC}\simeq9$--12, indicating ``strong'' to ``very strong'' evidence in favor of the linked-pair model over the baseline continuum model, according to the scale of \citet{burnham02}.
For comparison, a single broad Gaussian produces a smaller improvement, $\Delta\mathcal{C}/\Delta\mathrm{dof}\simeq12$--15/3, and provides a poorer description of the residual structure, favoring the two-line interpretation. As the same hierarchy of fit improvements is recovered for all three baseline continuum models, we conclude that the phenomenological evidence for the primary pair is not driven by the choice of continuum model.

To complement these statistical assessments, we performed Monte Carlo (MC) simulations to estimate the detection significance of the primary absorption feature. Specifically, we computed the probability that random fluctuations in the real spectra could produce absorption features as significant as those observed. Given the similar results obtained with the different baseline continuum models, we adopted \textsc{uxclumpy} due to its lower computational cost. We generated $2.5\times10^4$ \texttt{fakeit} realizations with no absorber, using the actual exposure times and response matrices of \xmm, \xrism, and \nustar data. Each simulated spectrum was then fitted following the same procedure applied to the real data, searching the line centroid with XSPEC \texttt{steppar} across 3.5--7~keV in steps of 5~eV. We restricted the search to 7~keV for computational efficiency and because the blind scan did not reveal significant features at higher energies in the real data. 
For the modeling, we considered the fiducial paired-line model described above, i.e., the Gaussian pair with $\sigma=10$~eV and a fixed centroid separation of $\Delta E_\mathrm{rest}=0.266$~keV. For each simulated spectrum, we computed the improvement in fit statistic relative to the model without the additional absorption component. The chance probability is then given by the fraction of simulations in which random noise produces a $\Delta\mathcal{C}$ equal to or larger than that measured in the real data \citep[e.g.,][]{markowitz06,vignali15,costanzo22,peca23axis}. For the primary pair, the simulations give an empirical confidence level of $99.98\%$, corresponding to a significance of $3.7\sigma$.

Finally, we verified that these results are recovered using the \resolve data alone, providing a consistency check on the joint analysis. Despite contributing only $\sim15\%$ of the counts used in the joint fit, \resolve yields absorber parameters, line centroids, and equivalent widths consistent with the joint-fit values, albeit with larger uncertainties. It also resolves the two absorption features at the expected Fe\,\textsc{xxv} He$\alpha$/Fe\,\textsc{xxvi} Ly$\alpha$ separation, showing that it is the main driver of the two-line solution. The linked Gaussian pair improves the fits by $\Delta\mathcal{C}/\Delta$dof $\simeq 10$--$11/3$ (2.4--2.5$\sigma$), with $\Delta\mathrm{AIC} \simeq 4$--5 relative to the baseline continua, while MC simulations yield a significance of 2.1$\sigma$, lower than in the joint
analysis because of the reduced photon statistics. Details of the \resolve-only fits and blind line scan are given in Appendix~\ref{app:resolve}.

\begin{deluxetable*}{lccccccccc}
\tabletypesize{\small}
\tablecaption{Statistical comparison of the best-fit absorption models relative to the corresponding baseline continua. The double-Gaussian model has $\sigma=10$~eV and line energies linked through a common absorber redshift, preserving $\Delta E_\mathrm{rest}=0.266$~keV. The \textsc{xstar} results use $v_\mathrm{turb}=1000$~km~s$^{-1}$. For each model, we list $\Delta\mathcal{C}$/$\Delta$dof, Gaussian-equivalent significance, and $\Delta$AIC with its evidence score. We follow the evidence scale of \citet{burnham02}: $4<\Delta\mathrm{AIC}<7$ indicates ``positive'' (P), $7<\Delta\mathrm{AIC}<10$ ``strong'' (S), and $\Delta\mathrm{AIC}>10$ ``very strong'' (VS) evidence in favor of the model with the lower AIC.
\label{tab:modelcomp}}
\tablewidth{0pt}
\tablehead{
\colhead{Model} &
\multicolumn{3}{c}{\textsc{MYTorus}} &
\multicolumn{3}{c}{\textsc{X-skirtor}} &
\multicolumn{3}{c}{\textsc{uxclumpy}} \\
\colhead{} &
\colhead{$\Delta\mathcal{C}$/$\Delta$dof} & \colhead{Sig.} & \colhead{$\Delta$AIC} &
\colhead{$\Delta\mathcal{C}$/$\Delta$dof} & \colhead{Sig.} & \colhead{$\Delta$AIC} &
\colhead{$\Delta\mathcal{C}$/$\Delta$dof} & \colhead{Sig.} & \colhead{$\Delta$AIC}
}
\startdata
$+$\,2 Gauss                    & $16.4/3$ & $3.3\sigma$ & $10.4$ / VS & $14.7/3$ & $3.1\sigma$ & $8.7$ / S & $17.5/3$ & $3.5\sigma$ & $11.5$ / VS \\
$+$\,\textsc{xstar}& $15.7/3$ & $3.2\sigma$ & $9.7$ / S & $13.3/3$ & $2.9\sigma$ & $7.3$ / S   & $17.1/3$ & $3.4\sigma$ & $11.1$ / VS \\
$+$\,\textsc{xstar} $3\,Z_\odot$  & $15.5/3$ & $3.2\sigma$ & $9.5$ / S & $13.2/3$ & $2.9\sigma$ & $7.2$ / S   & $16.7/3$ & $3.4\sigma$ & $10.7$ / VS \\
\enddata
\end{deluxetable*}

\subsection{Physical modeling} \label{sec:xstar}

We replaced the phenomenological Gaussian pair with a physically motivated photoionization absorber model, using a grid generated with \textsc{xstar}~v2.59 \citep{xstar1,xstar2}. Following \citetalias{peca25a} (and references therein) we assumed a power-law ionizing continuum with $\Gamma = 2$, an initial gas temperature of $T = 10^6$~K, a gas density of $n = 10^{10}$~cm$^{-3}$, a fixed covering factor of unity, and solar abundances \citep{asplund09}.
All \textsc{xstar} grids used in this work were computed at sub-eV spectral resolution up to the Fe K$\alpha$ region. The free parameters of the \textsc{xstar} grid are the column density of the absorber ($N_\mathrm{H,abs}$), the ionization parameter ($\log\xi$), the turbulent velocity ($v_\mathrm{turb}$), and the absorber redshift ($z_\mathrm{abs}$). The ionization parameter, expressed in units of erg~s$^{-1}$~cm, is defined as $\xi = L_\mathrm{ion}/(n\,r^2)$, where $L_\mathrm{ion}$ is the ionizing luminosity integrated between 1 and 1000~Ryd \citep{tarter69}. 
We also verified that adopting a harder ionizing continuum with $\Gamma=1.7$, representative of the average best-fitting slope across the three baseline models when including the absorption, yields absorber properties consistent within their uncertainties (Appendix~\ref{app:stability}).
The best-fit parameters for each baseline continuum model are reported in Table~\ref{tab:xstar}.

\subsubsection{Single-zone ionized absorber: the primary UFI} \label{sec:singlezone}

We initially allowed the turbulent velocity to vary, but it remained unconstrained. Indeed, for all baseline continuum models the fit returned only an upper limit of $v_\mathrm{turb}\lesssim5000$~km~s$^{-1}$ when fitting the primary pair. We therefore tested grids with fixed $v_\mathrm{turb}=100$, 1000, and 5000~km~s$^{-1}$. Among these, $v_\mathrm{turb}=1000$~km~s$^{-1}$ provides the best or statistically comparable fit across all baseline models and is favored by AIC over leaving $v_\mathrm{turb}$ free to vary. Therefore, we adopted $v_\mathrm{turb}=1000$~km~s$^{-1}$ as our fiducial choice, with results for the other tested turbulent-velocity grids reported in Appendix~\ref{app:fullgrid}.

\begin{figure*}[ht]
    \centering
    \includegraphics[width=\textwidth]{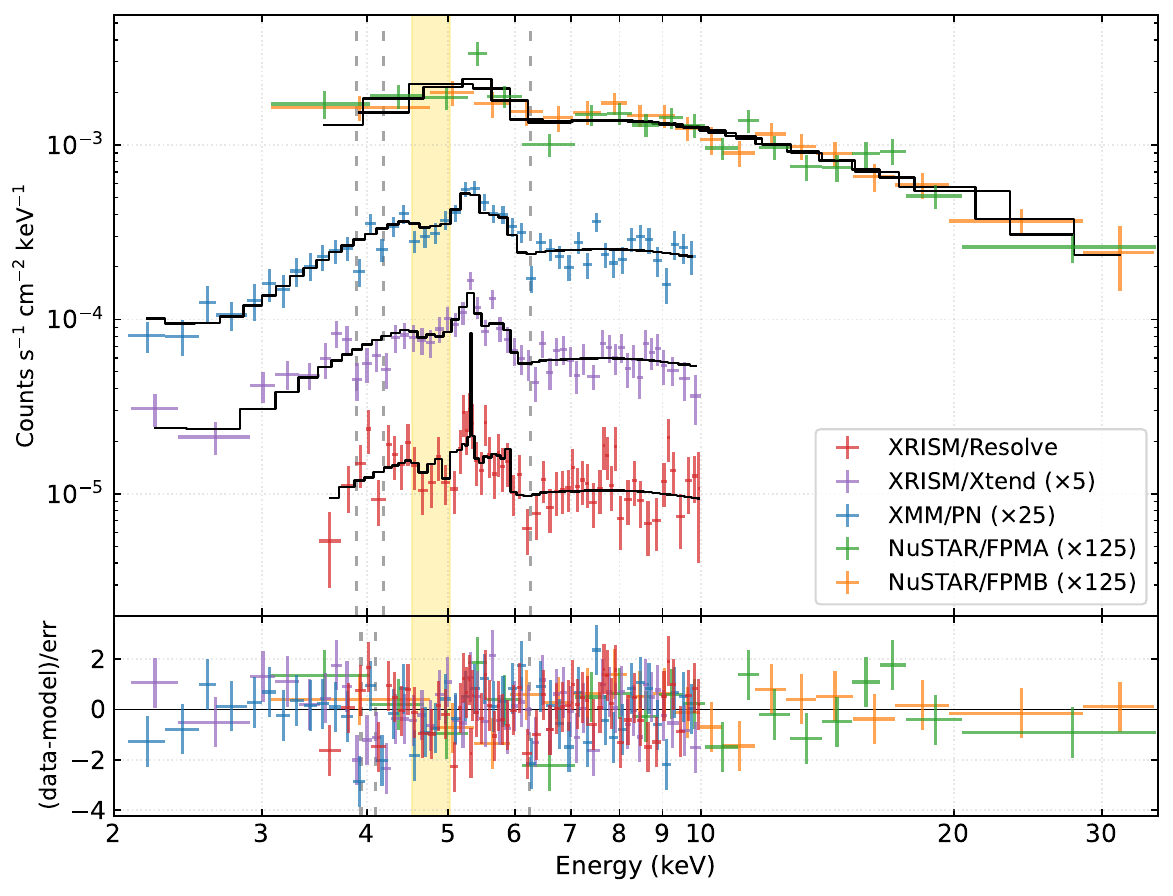}
    \caption{Best-fit observed-frame spectrum of ESP~39607 modeled with \textsc{MYTorus} and a solar-abundance, single-zone \textsc{xstar} absorber with $v_\mathrm{turb}=1000$~km~s$^{-1}$ (details in Section~\ref{sec:spectral}). Spectra and color-coding are as in Figure~\ref{fig:noabs}, with all spectra except \resolve\ shifted in normalization for visual clarity. The lower panel shows the residuals. The gold band marks the primary UFI at $v_\mathrm{in}\simeq0.16c$, modeled by Fe\,\textsc{xxv} He$\alpha$ and Fe\,\textsc{xxvi} Ly$\alpha$ absorption. Gray dashed lines mark the candidate secondary UFI near $\sim4$~keV and candidate UFO near 6.26~keV (see Section~\ref{sec:secondary}).}
    \label{fig:bestfit}
\end{figure*}

The \textsc{xstar} modeling reproduces the phenomenological two-Gaussian solution, with a single redshifted absorber naturally accounting for both the Fe\,\textsc{xxv} He$\alpha$ and Fe\,\textsc{xxvi} Ly$\alpha$ absorption lines (Figures~\ref{fig:bestfit} and~\ref{fig:resolve_zoom}). Although the grid includes absorption from all abundant ions, the observed structure is dominated by these two transitions.
The best-fit absorber redshift is $z_\mathrm{abs} \simeq 0.413$, consistent across all three baseline continuum models (Table~\ref{tab:xstar}). We estimated the inflow velocity using the relation
$(1+z_\mathrm{abs}) = (1+z_\mathrm{abs,int})(1+z_\mathrm{sys})$  \citep[e.g.,][]{tombesi11,bertola20,serafinelli23}, where $z_\mathrm{abs}$ is the observed redshift of the absorber, $z_\mathrm{abs,int}$ is the intrinsic absorber redshift, and $z_\mathrm{sys} = 0.201$ is the cosmological redshift of the source. 
This gives $z_\mathrm{abs,int} \simeq 0.177$. The intrinsic redshift is then related to the inflow velocity through the relativistic Doppler formula, $1+z_\mathrm{abs,int} = \sqrt{\frac{1+\beta}{1-\beta}}$, where $\beta = v_\mathrm{in}/c$.
This calculation yields an inflow velocity of $v_\mathrm{in}\simeq0.16c$, with an uncertainty of less than $0.01c$, identifying the absorber as a UFI. This value is in remarkable agreement with that reported by \citetalias{peca25a} from the earlier \nustar\ epochs. The absorber is further characterized by $\log N_\mathrm{H,abs}/\mathrm{cm}^{-2}\simeq23.7$--23.8 and $\log\xi/\mathrm{erg~s^{-1}~cm}\simeq3.7$--3.8, also consistent across all three baseline continuum models, as detailed in Table~\ref{tab:xstar}. 

This single-zone \textsc{xstar} model improves the fits across all three baseline continuum models. For $v_\mathrm{turb}=1000$~km~s$^{-1}$, the improvement is $\Delta\mathcal{C}/\Delta\mathrm{dof}\simeq13$--17/3, corresponding to Gaussian-equivalent significances of $2.9$--$3.4\sigma$. The \textsc{xstar} fits are also favored by the AIC, with $\Delta$AIC~$\simeq7$--11, indicating ``strong'' to ``very strong'' evidence relative to the baseline continuum models (Table~\ref{tab:modelcomp}). As in the phenomenological analysis, the recovery of the same absorber solution across all three models shows that the primary-UFI detection is not driven by a specific baseline continuum model. We note that, although the values remain consistent within their uncertainties, modeling the absorption with the \textsc{xstar} grid shifts $\Gamma$ to slightly higher, more canonical values \citep[e.g.,][]{ricci17,serafinelli17} and modestly reduces the scattering fraction, suggesting that the absorption residuals are better described by the absorber rather than partly compensated for by the continuum.

\begin{deluxetable*}{lccccccccc}
\tabletypesize{\small}
\tablecaption{Best-fit absorber and continuum parameters from the single-zone \textsc{xstar} models with $v_\mathrm{turb}=1000$~km~s$^{-1}$. From left to right: baseline continuum model; absorber redshift; absorber column density in units of $\log$~cm$^{-2}$; ionization parameter in units of erg~s$^{-1}$~cm in log scale; line-of-sight obscuring column density in units of $10^{22}$~cm$^{-2}$; photon index; scattered fraction; intrinsic rest-frame 2--10~keV luminosity in units of erg~s$^{-1}$ in log scale; fit statistic over degrees of freedom; and AIC value.
\label{tab:xstar}}
\tablewidth{0pt}
\tablehead{
\colhead{Model} & \colhead{$z_\mathrm{abs}$} & \colhead{$\log N_\mathrm{H,abs}$} & \colhead{$\log\xi$} & \colhead{$N_\mathrm{H}$} & \colhead{$\Gamma$} & \colhead{$f_\mathrm{scat}$ (\%)} & \colhead{$\log L_{2\text{--}10}^\mathrm{intr}$} & \colhead{$\mathcal{C}$/dof} & \colhead{AIC}
}
\startdata
\textsc{MYTorus}   & $0.413^{+0.005}_{-0.004}$ & $23.8^{+0.3}_{-0.7}$ & $3.8^{+0.2}_{-0.3}$ & $36.0^{+4.0}_{-3.3}$ & $1.65^{+0.18}_{-0.13}$ & $2.3^{+0.4}_{-0.7}$ & $44.54^{+0.21}_{-0.19}$ & 1004.4/1246 & 1026.4 \\
\textsc{X-skirtor}  & $0.413^{+0.007}_{-0.005}$ & $23.7^{+0.3}_{-0.8}$ & $3.7^{+0.2}_{-0.4}$ & $36.7^{+2.3}_{-2.9}$ & $1.59^{+0.12}_{-0.10}$ & $2.2^{+0.6}_{-0.5}$ & $44.52^{+0.19}_{-0.16}$ & 1006.3/1247 & 1026.3 \\
\textsc{uxclumpy} & $0.413\pm0.005$ & $23.8^{+0.3}_{-0.6}$ & $3.7^{+0.2}_{-0.3}$ & $38.0^{+4.7}_{-3.2}$ & $1.79^{+0.17}_{-0.09}$ & $3.6\pm1.0$ & $44.62^{+0.34}_{-0.14}$ & 1014.6/1247 & 1034.6 \\
\hline
\textsc{MYTorus} ($3\,Z_\odot$)  & $0.413\pm0.005$ & $23.3^{+0.3}_{-0.6}$ & $3.8^{+0.2}_{-0.3}$ & $35.9^{+4.1}_{-3.4}$ & $1.65^{+0.18}_{-0.13}$ & $2.3^{+0.7}_{-0.7}$ & $44.53^{+0.24}_{-0.16}$ & 1004.7/1246 & 1026.7 \\
\textsc{X-skirtor} ($3\,Z_\odot$)   & $0.413^{+0.007}_{-0.005}$ & $23.2^{+0.3}_{-0.8}$ & $3.7^{+0.2}_{-0.4}$ & $36.7^{+2.4}_{-2.9}$ & $1.59^{+0.12}_{-0.10}$ & $2.2^{+0.7}_{-0.5}$ & $44.52^{+0.14}_{-0.14}$ & 1006.4/1247 & 1026.4 \\
\textsc{uxclumpy} ($3\,Z_\odot$) & $0.413\pm0.005$ & $23.3^{+0.3}_{-0.6}$ & $3.8^{+0.2}_{-0.3}$ & $38.0^{+4.8}_{-3.3}$ & $1.79^{+0.17}_{-0.10}$ & $3.6^{+1.1}_{-1.0}$ & $44.63^{+0.21}_{-0.14}$ & 1014.9/1247 & 1034.9
\enddata
\end{deluxetable*}

\subsubsection{Multiphase inflow at the primary UFI velocity} \label{sec:multiphase}

For completeness, we also tested whether the data support a multiphase, stratified absorber at the same velocity as the primary UFI, a configuration commonly invoked for ionized AGN winds \citep[e.g.,][]{lanzuisi24,xrism25_multiufo,reeves26}. Starting from the best-fit single-zone solution ($z_\mathrm{abs}\simeq0.413$, $v_\mathrm{in}\simeq0.16c$, $\log N_\mathrm{H,abs}/\mathrm{cm}^{-2}\simeq23.7$--23.8, and $\log\xi/\mathrm{erg~s^{-1}~cm}\simeq3.7$--3.8 for the solar grid with $v_\mathrm{turb}=1000$~km~s$^{-1}$), we added a second \textsc{xstar} component with its redshift tied to that of the primary zone and the same turbulent velocity, leaving its column density and ionization parameters free to vary.

The fit converges to a lower-column, lower-ionization phase coexisting with a higher-column, higher-ionization phase. However, the improvement is not significant relative to the single-zone model: across the three baseline continua, the same-redshift two-zone model gives $\Delta\mathcal{C}/\Delta\mathrm{dof}\simeq(3.2$--$3.9)/2$ and no support by AIC (Appendix~\ref{app:fullgrid}, Table~\ref{tab:modelcomp_full}). We conclude that the current data do not require an additional ionization phase at the primary UFI velocity, and that the current signal-to-noise prevents a more detailed exploration of the possible ionization stratification of the inflow. 

\subsubsection{Metallicity dependence} \label{sec:metallicity}

\begin{figure*}[ht]
    \centering
        \includegraphics[width=0.75\textwidth]{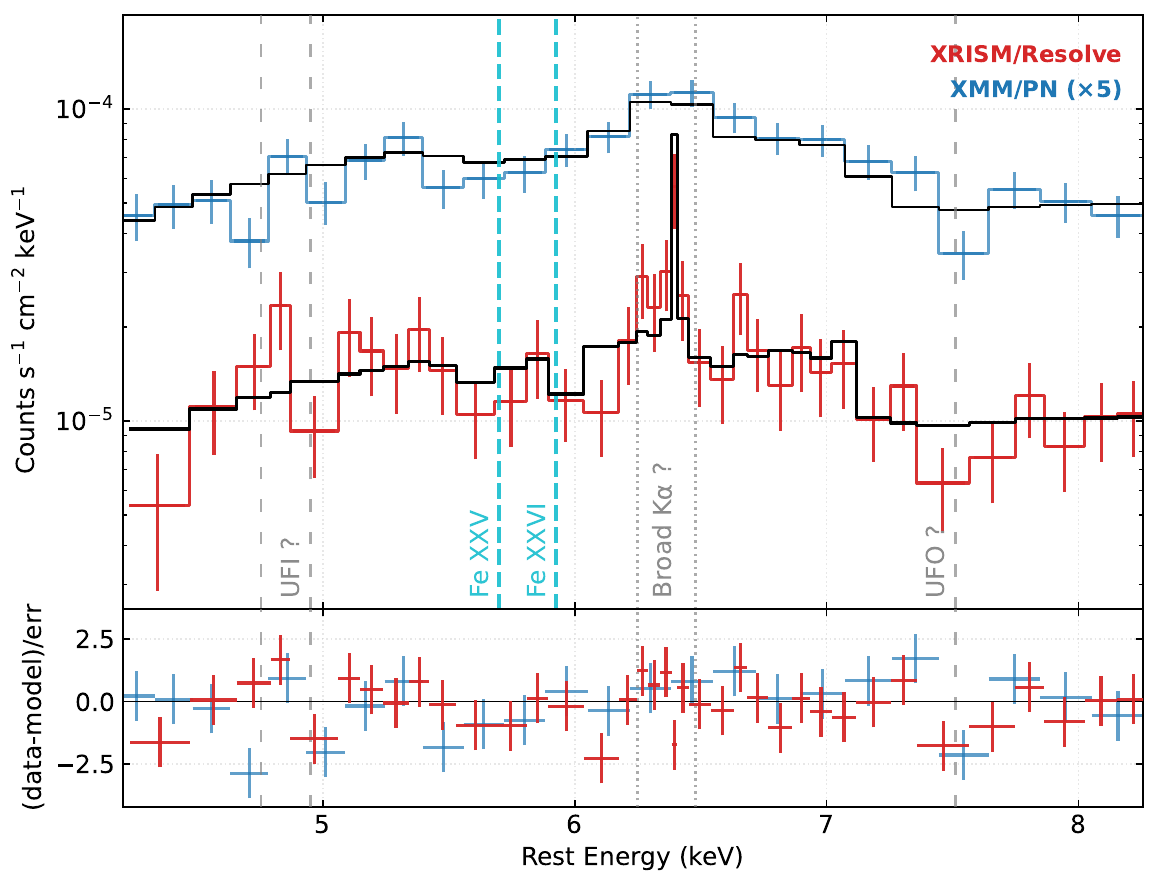}
    \caption{Zoomed-in view of the \xrism/\resolve (red) and \xmm/PN (blue) spectra of ESP~39607, shown in the source rest frame, with the PN normalization multiplied by 5 for visual clarity. The black solid curve shows the \textsc{MYTorus} plus solar-abundance, single-zone \textsc{xstar} model. Cyan dashed lines mark the Fe\,\textsc{xxv} He$\alpha$ and Fe\,\textsc{xxvi} Ly$\alpha$ transitions of the primary UFI at $\sim$5.6 and $\sim$5.9~keV, corresponding to $v_\mathrm{in}\simeq0.16c$. Gray dashed lines indicate the candidate secondary UFI at $\sim$4.8 and $\sim$5.0~keV ($v_\mathrm{in}\sim0.31c$) and the candidate UFO near $\sim$7.5~keV (Section~\ref{sec:secondary}), while the gray dotted lines bracket the broad Fe~K$\alpha$ emission region (Section~\ref{sec:feka}). The lower panel shows the residuals. Data are rebinned for graphic purposes. 
    \label{fig:resolve_zoom}}
\end{figure*}

To test the robustness of the primary UFI detection against the assumed iron abundance, and motivated by recent \xrism\ measurements of super-solar nickel-to-iron abundances in obscured AGN \citep[e.g.,][]{xrism_supsolar,ngc1068_xrism}, we computed an additional \textsc{xstar} grid with $3\,Z_\odot$ abundances.
For $v_\mathrm{turb}=1000$~km~s$^{-1}$, the $3\,Z_\odot$ grid improves the fits relative to the baseline continuum models by $\Delta\mathcal{C}/\Delta\mathrm{dof}\simeq13$--17/3, corresponding to significances of $2.9$--$3.4\sigma$. The models are also favored by the AIC, with $\Delta\mathrm{AIC}\simeq7$--11 (Table~\ref{tab:modelcomp}). The two abundance grids are statistically indistinguishable, with the fit statistic differing by only $\Delta\mathcal{C}<0.3$ across the three baseline continuum models, and identical numbers of free parameters, so the AIC differences are the same.
The absorber column density follows the expected metallicity scaling, decreasing to $\log N_\mathrm{H,abs}/\mathrm{cm}^{-2}\simeq23.2$--23.3 from $\simeq23.7$--23.8 for the solar-abundance grid. This reflects the lower hydrogen-equivalent column required to produce the same Fe~K opacity at higher iron abundance. The other absorber parameters remain essentially unchanged, with $z_\mathrm{abs}\simeq0.413$ and $\log\xi/\mathrm{erg~s^{-1}~cm}\simeq3.7$--3.8. At the present signal-to-noise ratio, however, the solar and $3\,Z_\odot$ model curves are nearly indistinguishable. We therefore show only the solar-abundance model in Figure~\ref{fig:resolve_zoom}, which presents a zoomed view of the \resolve and PN spectra in the source rest frame.

\begin{figure*}[ht]
    \centering
    \includegraphics[width=0.99\textwidth]{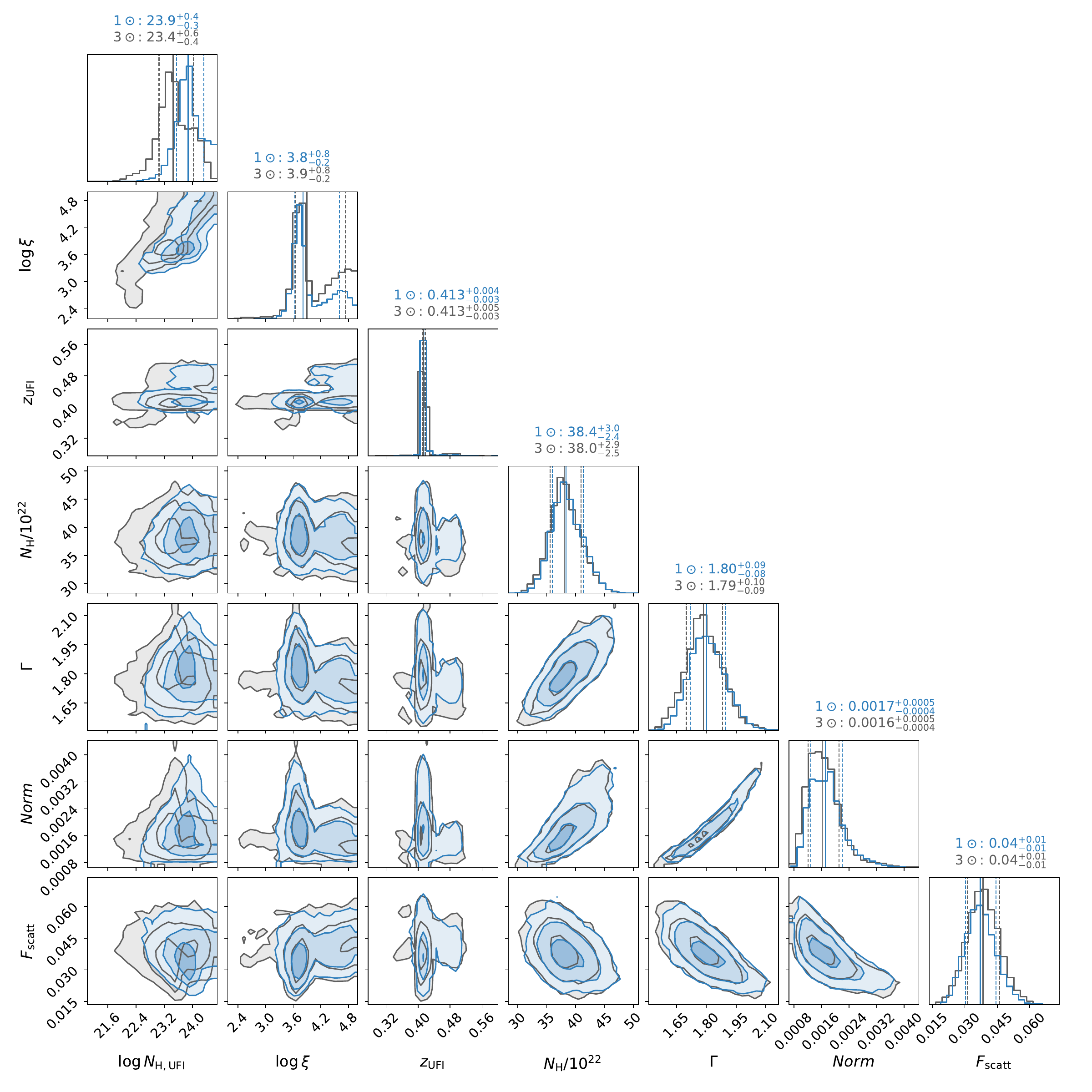}
    \caption{Posterior distributions of the \textsc{uxclumpy} + single-zone \textsc{xstar} model parameters for the solar (blue) and $3\,Z_\odot$ (gray) metallicity grids ($v_\mathrm{turb} = 1000$~km~s$^{-1}$; Section~\ref{sec:metallicity}). Off-diagonal panels show 2D marginal posteriors at 1, 2, and $3\sigma$ confidence contours. Diagonal panels show 1D posteriors, with values above each histogram showing the medians with 16th and 84th percentile uncertainties. From left to right, the panels show the logarithm of the absorber column density in cm$^{-2}$, the logarithm of the ionization parameter in erg~s$^{-1}$~cm, the absorber redshift, the AGN line-of-sight column density in units of $10^{22}$ cm$^{-2}$, the photon index, the power-law normalization in photons~keV$^{-1}$~cm$^{-2}$~s$^{-1}$, and the scattered fraction. The MCMC medians are consistent with the best-fit values reported in Table~\ref{tab:xstar}.}
    \label{fig:corner_ufi}
\end{figure*}

We further explored the \textsc{xstar} solutions for both metallicity grids using Markov chain Monte Carlo (MCMC) simulations within XSPEC, adopting \textsc{uxclumpy} as for the MC simulations of Section~\ref{sec:gaussians}. We used 50 walkers, a chain length of 500,000 steps, and a burn-in phase of 10\%. Chain convergence was checked visually from the post-burn-in traces and with the Geweke diagnostic \citep{geweke92}, finding all parameters consistent with convergence ($|z|<2$). The posteriors are shown in Figure~\ref{fig:corner_ufi}. The chains show that the continuum posteriors are essentially unchanged between the two metallicities, and that the ionization parameter is likewise consistent, both posteriors sharing the same shape with a principal peak and an extended high-ionization tail. The main absorber difference is the expected shift in hydrogen-equivalent column density. The MCMC medians and 16th--84th percentile intervals for the absorber parameters are $\log N_\mathrm{H,abs}/\mathrm{cm}^{-2}=23.9^{+0.4}_{-0.3}$, $\log\xi/\mathrm{erg~s^{-1}~cm}=3.8^{+0.8}_{-0.2}$, and $z_\mathrm{abs}=0.413^{+0.004}_{-0.003}$ for the solar grid, and $23.4^{+0.6}_{-0.4}$, $3.9^{+0.8}_{-0.2}$, and $0.413^{+0.005}_{-0.003}$ for the $3\,Z_\odot$ grid, respectively. The continuum parameters are likewise consistent between the two grids: $N_\mathrm{H}=38.4^{+3.0}_{-2.4}\times10^{22}$~cm$^{-2}$, $\Gamma=1.80^{+0.09}_{-0.08}$, and $f_\mathrm{scat}=4\pm1\%$ for the solar grid, compared with $38.0^{+2.9}_{-2.5}\times10^{22}$~cm$^{-2}$, $1.79^{+0.10}_{-0.09}$, and $f_\mathrm{scat}=4\pm1\%$ for the $3\,Z_\odot$ grid. These posterior medians are consistent with the XSPEC best-fit values in Table~\ref{tab:xstar} within the corresponding MCMC intervals for all fitted parameters.

\subsection{Secondary components} \label{sec:secondary}

In addition to the primary UFI, the residuals show two weaker absorption structures that we model separately: a lower-energy pair at observed-frame $\sim$3.9 and $\sim$4.1~keV and a single residual at $\sim$6.26~keV. We interpret the former as a candidate secondary UFI and the latter as a candidate UFO. We test both the phenomenological descriptions and the physically motivated \textsc{xstar} absorbers against all three baseline continuum models, as done for the primary UFI. The full per-model results are reported in Appendix~\ref{app:fullgrid}, Table~\ref{tab:modelcomp_full}.

\subsubsection{The candidate secondary UFI} \label{sec:secondary_ufi}

We first characterized the secondary pair phenomenologically with two narrow Gaussian absorption lines. Allowing the widths to vary left them unconstrained, so we fixed both at $\sigma=10$~eV, as done for the primary UFI. With both centroids free, the model improves the baseline continua by $\Delta\mathcal{C}/\Delta\mathrm{dof}\simeq14$--17/4, whereas the linked-energy pair gives $\simeq10$--11/3. Across these comparisons, the Gaussian-equivalent significances span $2.4$--$3.1\sigma$ and the AIC improvements span $\Delta\mathrm{AIC}\simeq4$--9. As for the primary UFI, we adopted the linked-energy model as the fiducial phenomenological description because it enforces the expected Fe\,\textsc{xxv} He$\alpha$/Fe\,\textsc{xxvi} Ly$\alpha$ spacing and avoids over-interpreting the individual centroids. A single broad absorption Gaussian did not converge to a stable solution.

We then modeled the pair with a single \textsc{xstar} absorber, which naturally accounts for both lines as Fe\,\textsc{xxv} He$\alpha$ and Fe\,\textsc{xxvi} Ly$\alpha$ absorption from one ionized zone. The grids favor $v_\mathrm{turb}=1000$~km~s$^{-1}$ across the three baseline continua, the same value preferred by the primary UFI. This solution gives $z_\mathrm{abs}\simeq0.66$, corresponding to $v_\mathrm{in}\sim0.31c$, and improves the fits by $\Delta\mathcal{C}/\Delta\mathrm{dof}\simeq12$--16/3, corresponding to $2.7$--$3.3\sigma$ and $\Delta\mathrm{AIC}\simeq6$--10. When the secondary \textsc{xstar} zone is instead added to models already including the primary-UFI \textsc{xstar} absorber, the improvement across the three baseline continua is $\Delta\mathcal{C}/\Delta\mathrm{dof}=7.7$--$10.1/3$, corresponding to $1.9$--$2.4\sigma$, with $\Delta\mathrm{AIC}=1.7$--$4.1$.

Several considerations nevertheless argue for caution. The relevant 3.8--4.2~keV \resolve band has a lower signal-to-noise ratio than the primary-UFI band. Furthermore, the higher-order Fe K transitions of the putative second \textsc{xstar} zone, primarily Fe\,\textsc{xxv} He$\beta$ and Fe\,\textsc{xxvi} Ly$\beta$, are redshifted to $\sim4.8$ and $\sim5.0$~keV, respectively, partially overlapping the primary UFI pair and potentially inflating the single-zone-on-baseline significance. The more conservative two-zone result above, in which the primary UFI is already modeled, therefore provides the more appropriate measure of the additional component. As an empirical calibration of the feature strength, we applied the same MC simulations used for the primary UFI to the fixed-width, linked-energy Gaussian description of the secondary pair. In the observed spectra, the model improvement is $\Delta\mathcal{C}/\Delta\mathrm{dof}=10.2/3$ relative to the baseline \textsc{uxclumpy} continuum. The simulations yield a confidence level of $99.53\%$, or $2.8\sigma$. We therefore regard the secondary UFI solution as statistically interesting, although still tentative.

\subsubsection{The candidate UFO} \label{sec:ufo}

We then modeled the absorption residual at $\sim6.26$~keV with a narrow Gaussian. We fixed the width at $\sigma=10$~eV because it remained unconstrained when left free and verified that this choice does not affect the results. Relative to the baseline continua, the fits improved by $\Delta\mathcal{C}/\Delta\mathrm{dof}\simeq5$--7/2 across the three continuum models, corresponding to $1.9$--$2.2\sigma$, while the AIC improves only marginally, by $\Delta\mathrm{AIC}\simeq1$--3. 
We then applied the same MC framework used for the primary UFI, but with a single fixed-width Gaussian line on top of the baseline continuum. This gives a confidence level of $95.05\%$, corresponding to $2.0\sigma$. When the same line is added to the fiducial solar-abundance \textsc{xstar} fits for the primary UFI, it improves the fits by $\Delta\mathcal{C}/\Delta\mathrm{dof}=4.5$--$6.0/2$ across the three baseline continua, with $\Delta\mathrm{AIC}=0.5$--$2.0$, again indicating only marginal AIC support. As for the primary and secondary pairs, we assume iron as the most likely species, given its high cosmic abundance. If identified as blueshifted Fe\,\textsc{xxvi} Ly$\alpha$, the implied outflow velocity is $v_\mathrm{out}\sim 0.08c$, while for Fe\,\textsc{xxv} He$\alpha$ it is $\sim0.12c$, placing the feature in the UFO velocity regime under either identification.

We also explored whether the feature could be described self-consistently with an outflowing \textsc{xstar} absorber, but the limited signal-to-noise prevented the absorber properties from being meaningfully constrained. We therefore regard the UFO identification as tentative.

\subsection{The 6.4 keV \texorpdfstring{Fe K$\alpha$}{Fe K-alpha} emission profile} \label{sec:feka}

The residuals in the \resolve\ spectrum suggest that the neutral Fe~K$\alpha$ emission near 6.4~keV might not be fully described by the narrow component included in the torus models. We therefore added a Gaussian emission line at 6.4~keV in the source rest frame, with its width free to vary. In the baseline models, this component improves the \textsc{MYTorus} and \textsc{uxclumpy} fits by $\Delta\mathcal{C}/\Delta\mathrm{dof}=9.3$--$10.4/2$ ($2.6$--$2.8\sigma$), with $\Delta\mathrm{AIC}=5.3$--$6.4$. The line width converges to $\sigma\sim60$--80~eV, corresponding to a full width at half maximum (FWHM) of $\sim7000$--9000~km~s$^{-1}$, although it remains weakly constrained. The improvement is weaker for \textsc{X-skirtor}, at $\Delta\mathcal{C}/\Delta\mathrm{dof}=5.1/2$ ($1.8\sigma$), with $\Delta\mathrm{AIC}=1.1$. 
The same model dependence remains after including the primary UFI absorber, with the broad Gaussian improving the \textsc{MYTorus} and \textsc{uxclumpy} fits by $\Delta\mathcal{C}/\Delta\mathrm{dof}=7.0$--$8.5/2$ ($2.2$--$2.5\sigma$), with $\Delta\mathrm{AIC}=3.0$--$4.5$, and a line width consistent with the baseline range. \textsc{X-skirtor} again shows a weaker improvement, with $\Delta\mathcal{C}/\Delta\mathrm{dof}=3.3/2$ ($1.3\sigma$) and $\Delta\mathrm{AIC}=-0.7$, which provides no support for the additional component.
In all cases, we verified that including or excluding the broad component does not change the derived source properties or the primary UFI parameters, nor does it affect the UFI identification.

The differing behavior across the torus models likely reflects their treatment of the reflected continuum, the scattering geometry, and the implementation of the Fe~K$\alpha$ complex. In particular, the \textsc{uxclumpy} tables model Fe~K$\alpha$ as a single unresolved line rather than including the intrinsic K$\alpha_1$/K$\alpha_2$ doublet structure present in \textsc{MYTorus} and \textsc{X-skirtor}. Part of the residual at the \resolve\ resolution may therefore arise from this simplified treatment. We consequently regard the broad component as tentative.

To investigate the line profile more directly, we used \textsc{MYTorus}, whose fluorescent-line module can be broadened or switched off independently of the continuum. We first tested whether the residuals could be accounted for by broadening the narrow-line module with \textsc{gsmooth}. The fit improves by only $\Delta\mathcal{C}/\Delta\mathrm{dof}=0.6/1$ and is disfavored by the AIC, with $\Delta\mathrm{AIC}=-1.4$. We then switched off the fluorescent line module and modeled the 6.4~keV Fe~K$\alpha$ emission with two Gaussian components whose widths were free to vary. The fit separates them into a narrow core and a broader component. The narrow core, $\sigma\simeq5\pm4$~eV, is essentially unresolved ($\lesssim1000$~km~s$^{-1}$) and consistent with the narrow, distant-torus Fe~K$\alpha$ core resolved by \xrism\ in both obscured \citep{ngc1068_xrism,ngc4388_xrism} and unobscured \citep{dadina26,kammoun25,li26} AGN. The broad component, with $\sigma\simeq69\pm36$~eV (FWHM $\sim7600$~km~s$^{-1}$), is consistent with the width inferred above by adding a broad Gaussian to the narrow line already included in the torus models. It is also broadly consistent with the intermediate-width neutral Fe~K$\alpha$ emission attributed to the broad-line region or the inner torus wall in other Seyfert galaxies \citep[e.g.,][]{xrism24_feka,ngc4388_xrism,li26}. We emphasize that we used this decomposition only to characterize the line widths, rather than as the preferred spectral model, because the AIC does not favor it over the standard \textsc{MYTorus} line module with an additional broad Gaussian.

\section{Discussion} \label{sec:discussion}

The simultaneous \xrism, \xmm, and \nustar\ campaign combines the high spectral resolution of \resolve\ with the broadband photon statistics of the full dataset. The \resolve\ spectrum shows that the redshifted Fe~K absorption feature previously detected with \nustar\ contains two narrow components consistent with Fe\,\textsc{xxv} He$\alpha$ and Fe\,\textsc{xxvi} Ly$\alpha$ absorption from a single absorber at $v_\mathrm{in}\simeq0.16c$. The feature is recovered in joint fits to the full dataset across all three continuum prescriptions and is described consistently by the phenomenological and photoionization models. We now examine the recurrence of this feature and its physical implications.

\subsection{Multi-epoch comparison} \label{sec:comparison}

The UFI velocity measured in this work, $v_\mathrm{in}\simeq0.16c$, agrees with that reported by \citetalias{peca25a} for the May~2023 and August~2024 \nustar\ observations, which together with the 2026 campaign span 2.2~yr in the source rest frame. Its recovery in an independent third epoch, with the previously blended feature now resolved into the Fe\,\textsc{xxv} He$\alpha$/Fe\,\textsc{xxvi} Ly$\alpha$ pair thanks to the high spectral resolution of \resolve, substantially strengthens the UFI interpretation.
The other \textsc{xstar} parameters are likewise compatible with the earlier analysis within the uncertainties. \citetalias{peca25a} adopted $v_\mathrm{turb}=5000$~km~s$^{-1}$ as their fiducial value, while also finding a statistically acceptable solution with $v_\mathrm{turb}=1000$~km~s$^{-1}$. At the $\sim400$~eV resolution of \nustar, the two Fe~K transitions are unresolved and appear as a single absorption feature, preventing a clear distinction between the turbulent-velocity grids and leaving the ionization only weakly constrained. By resolving the two transitions, \resolve\ instead favors $v_\mathrm{turb}=1000$~km~s$^{-1}$, without introducing tension with the previous \nustar\ results. Taken together, these measurements indicate similar absorber properties across the three observations within the current uncertainties.

Because the MC analyses directly calibrate the probability of obtaining a feature-like improvement under their respective no-feature null models, these probabilities can be combined across statistically independent datasets with Fisher's method \citep{mosteller_fisher48,zoghbi16}. Combining the joint significance of $4.2\sigma$ for the 2023--2024 observations \citepalias{peca25a} with the independent 2026 significance of $3.7\sigma$, both obtained with the same continuum model and MC procedure, gives an overall significance of $5.3\sigma$. This combination, however, assesses the evidence against the joint no-feature hypothesis and does not by itself establish that the same physical inflow produced all three detections.

To assess whether the three epochs could instead contain unrelated features at arbitrary energies, we estimate the chance probability of their observed energy alignment. We compare the observed-frame centroids of the blended \nustar\ feature in 2023 and 2024, at 4.8 and 4.5~keV, with the components resolved in the simultaneous 2026 fit at 4.67 and 4.92~keV (see Tables~\ref{tab:lines} and~\ref{tab:nustar_epochs} in Appendix~\ref{app:linechar_var}). Using the Fe\,\textsc{xxv} He$\alpha$ component gives a total span of $\Delta E=0.30$~keV, while Fe\,\textsc{xxvi} Ly$\alpha$ gives $\Delta E=0.42$~keV.
Assuming that unrelated redshifted Fe~K features are independently and uniformly distributed between $E_\mathrm{min}=3$~keV, the lower bound of the \nustar\ fitting band, and $E_\mathrm{max}$, the observed energy of the corresponding transition at the systemic redshift, which is 5.58~keV for Fe\,\textsc{xxv} He$\alpha$ and 5.80~keV for Fe\,\textsc{xxvi} Ly$\alpha$, the probability that three energies span no more than $\Delta E$ is
\begin{equation}
P_\mathrm{range}=3\left(\frac{\Delta E}{E_\mathrm{max}-E_\mathrm{min}}\right)^{2}-2\left(\frac{\Delta E}{E_\mathrm{max}-E_\mathrm{min}}\right)^{3}.
\end{equation}
This yields $P_\mathrm{range}=0.037$ and $0.061$ for the Fe\,\textsc{xxv} He$\alpha$ and Fe\,\textsc{xxvi} Ly$\alpha$ comparisons, respectively, corresponding to a chance probability of about 4--6\%. We stress that this is only an order-of-magnitude indication rather than a calibrated significance. Nevertheless, under this simplified null hypothesis, the probability of such an alignment is only a few percent, supporting the interpretation that the detections trace a recurrent inflow at a similar characteristic velocity rather than unrelated features. The physical implications of this recurrence are discussed in Section~\ref{sec:disc_cascade}.

Over the $\sim$15-year baseline spanned by the X-ray observations suitable for a detailed spectral analysis \citepalias{peca25a}, the continuum flux of ESP~39607 shows a long-term decrease. The earliest is the December~2010 \suzaku\ epoch, during which no UFI was detected (Appendix~\ref{app:suzaku}), when the source was $\sim$22\% brighter than in May~2023 and a factor of $\sim$2 brighter than in 2026. The steepest single-interval change occurred between the 2023 and 2024 \nustar\ epochs, with a $\sim$35\% drop in 15~months. The 2026 broadband fits from this work yield an observed 2--10~keV flux of $F_{2\text{--}10}^\mathrm{obs}\simeq9.3\times10^{-13}$~erg~s$^{-1}$~cm$^{-2}$, an additional $\sim$7\% decrease relative to 2024, for a cumulative decline of $\sim$50\%. The measured line-of-sight column densities remain mutually consistent across these epochs, within their uncertainties, suggesting that this decline is intrinsic rather than driven by increasing obscuration. The line equivalent widths nevertheless remain consistent within their uncertainties (Appendix~\ref{app:linechar_var}), with no evidence that the absorber properties changed as the continuum declined.

\subsection{Alternative interpretations} \label{sec:alt_ids}

Before quantifying the physical implications of the UFI, we assess whether the observed features could instead be explained by alternative line identifications or by gravitational redshift without bulk inflow.
First, we consider blueshifted transitions from elements lighter than iron. For the primary pair near $\sim$4.7 and 4.9~keV, this possibility was already examined by \citetalias{peca25a}, who considered the single, blended feature at $\sim$4.8~keV. They showed that identifications with Mg\,\textsc{xii}, Si\,\textsc{xiv}, or S\,\textsc{xvi} Ly$\alpha$ would require extreme velocities, $v_\mathrm{out}\sim0.7$--$0.9c$. Other species such as Ar\,\textsc{xviii}, K\,\textsc{xix}, and Ca\,\textsc{xx} Ly$\alpha$ would reduce the required velocity to $v_\mathrm{out}\sim0.3$--$0.5c$, but are more than an order of magnitude less abundant than Fe, with K nearly three orders of magnitude less abundant \citep[e.g.,][]{asplund09}. An interpretation in which the resolved pair is dominated by transitions from Ar, K, or Ca is therefore physically difficult to justify.

For the tentative secondary pair near $\sim$3.9 and 4.1~keV, the lower observed energy makes the identifications with elements lighter than Fe more plausible from a purely kinematic standpoint. Ar\,\textsc{xviii} and K\,\textsc{xix} Ly$\alpha$ would imply $v_\mathrm{out}\sim0.33$--$0.37c$ and $\sim0.23$--$0.28c$, respectively. A Ca\,\textsc{xix} He$\alpha$/Ca\,\textsc{xx} Ly$\alpha$ pair interpretation is the most favorable case, giving a common outflow velocity of $v_\mathrm{out}\sim0.18c$. Nevertheless, the abundance argument remains problematic for Ar, K, and Ca. By contrast, Mg\,\textsc{xii}, Si\,\textsc{xiv}, and S\,\textsc{xvi} Ly$\alpha$ would still require much larger velocities, $v_\mathrm{out}\sim0.5$--$0.8c$ respectively. We tested these alternatives by allowing the \textsc{xstar} absorber redshift to move into the outflow regime, but found no support for a blueshifted solution involving elements lighter than Fe. Therefore, while these interpretations cannot be fully ruled out based on the implied outflow velocities alone, we regard such lighter-element outflow interpretations as disfavored overall.

Second, we consider the possibility that the redshift of the absorption features arises from gravitational redshift in the strong-field region near the SMBH rather than from bulk infall \citep[e.g.,][]{nandra99,ruszkowski00,reeves05,yaqoob05}. Reproducing the primary pair through gravitational redshift alone would require absorbing material at $R\simeq7\,R_\mathrm{g}$, where the local orbital velocity reaches $\simeq0.44c$ and would be expected to imprint broad, smeared absorption. The observed lines are instead narrow (Section~\ref{sec:gaussians}), and a single broad-line description is not preferred over the two narrow components (Table~\ref{tab:modelcomp_full}). The required radius is even smaller for the tentative secondary pair, at $R\simeq4\,R_\mathrm{g}$, inside the innermost stable circular orbit at $6\,R_\mathrm{g}$ for a non-rotating black hole. Absent an additional supporting force, gas at these radii would fall into the black hole within hours, far shorter than the 2.2~yr source-rest-frame baseline over which the feature is observed to recur. We therefore conclude that, taken together, these considerations disfavor a purely gravitational origin.

\subsection{Inflow location and energetics} \label{sec:disc_energetics}

To test the physical consistency of the inflow velocity derived in Section~\ref{sec:singlezone}, we compare it with a simple model of radially infalling gas under gravitational attraction modulated by radiation pressure \citep[following][]{longinotti07}. Under this approximation, the characteristic radius at which the observed inflow velocity equals the radial free-fall velocity is
\begin{equation}\label{eq:rlaunch}
R = \frac{2\,G\,M_\mathrm{BH}}{v_\mathrm{in}^2}\,P_\mathrm{rad}\,,
\end{equation}
where $P_\mathrm{rad} = (1-L_\mathrm{bol}/L_\mathrm{Edd})$ is the radiation pressure term. We adopt $\log M_\mathrm{BH}/M_\odot = 8.26$ \citep{peca25b} and the average $\log L_{2\text{--}10}/\mathrm{erg~s^{-1}} \simeq 44.56$ across the three solar-abundance \textsc{xstar} continuum models (Table~\ref{tab:xstar}), which converts to $L_\mathrm{bol}\simeq8.26\times10^{45}$~erg~s$^{-1}$ and $L_\mathrm{bol}/L_\mathrm{Edd} \simeq 0.36$ using the \citet{duras20} bolometric correction. This yields $P_\mathrm{rad} \simeq 0.64$. Substituting it into Equation~(\ref{eq:rlaunch}) gives $R \simeq 49\,R_\mathrm{g}$, or $4.2\times10^{-4}$~pc. When radiation pressure is neglected ($P_\mathrm{rad} = 1$), the equation reduces to pure free-fall and yields $R \simeq 77\,R_\mathrm{g}$, or $6.7\times10^{-4}$~pc. Both values are comparable to those inferred for UFIs in the literature \citep[e.g.,][]{longinotti07,pounds18} and are consistent with the $R \simeq 22$--$89\,R_\mathrm{g}$ range derived by \citetalias{peca25a}.

At these radii, gravitational redshift makes only a small contribution to the observed shift. For a non-rotating black hole, gas at a distance $R$ produces a gravitational redshift $1+z_\mathrm{grav}=(1-2\,R_\mathrm{g}/R)^{-1/2}$, which at $R\simeq49$--$77\,R_\mathrm{g}$ amounts to $z_\mathrm{grav}\simeq1$--2\%. Although this correction exceeds the formal uncertainty on $z_\mathrm{abs}$, it only lowers the inferred velocity from $v_\mathrm{in}\simeq0.16c$ to $\simeq0.15c$. We therefore quote the directly measured Doppler velocity throughout. The alternative interpretation in which the entire shift is purely gravitational is disfavored for the reasons discussed in Section~\ref{sec:alt_ids}.

Following the thin-shell formalism commonly used to estimate the energetics of highly ionized AGN winds \citep[e.g.,][]{tombesi13,tombesi15,gofford15,serafinelli19} and adapting it to an inflow geometry, we estimate the mass inflow rate and kinetic luminosity as
\begin{equation}\label{eq:mdot}
\dot{M}_\mathrm{in} = 4\pi\,\mu\,m_\mathrm{p}\,N_\mathrm{H,abs}\,v_\mathrm{in}\,R\,C_f\,,
\end{equation}
\begin{equation}\label{eq:lkin}
L_\mathrm{kin} = \tfrac{1}{2}\,\dot{M}_\mathrm{in}\,v_\mathrm{in}^2\,,
\end{equation}
where $m_\mathrm{p}$ is the proton mass, $N_\mathrm{H,abs}$ is the column density of the inflowing material, $C_f$ is the global covering fraction, and $\mu=1.4$ is the mean atomic mass per proton. This expression corresponds to a shell-like geometry with characteristic thickness $\Delta R \sim R$ and gas density $n \sim N_\mathrm{H,abs}/R$. Since the covering fraction of the UFI gas is not constrained by the data, we assume $C_f=0.5$, by analogy with UFO studies where this value is commonly adopted \citep[e.g.,][]{tombesi10,serafinelli19}. Both $\dot{M}_\mathrm{in}$ and $L_\mathrm{kin}$ scale linearly with $\mu$, $N_\mathrm{H,abs}$, $R$, and $C_f$.

Using the characteristic radius across the two limits, $R \simeq 49$--$77\,R_\mathrm{g}$, we obtain $\dot{M}_\mathrm{in} \simeq 0.9$--$1.4\,M_\odot$~yr$^{-1}$ when using $N_\mathrm{H,abs}$ obtained with the solar metallicity \textsc{xstar} grid, and $\dot{M}_\mathrm{in} \simeq 0.27$--$0.42\,M_\odot$~yr$^{-1}$ when using the value from the $3\,Z_\odot$ grid, where fewer hydrogen atoms are needed to produce the same Fe~K opacity. 
The corresponding kinetic luminosities are $L_\mathrm{kin} \simeq (6.4$--$10)\times10^{44}$~erg~s$^{-1}$ for the solar-abundance case and $L_\mathrm{kin} \simeq (1.9$--$3.0)\times10^{44}$~erg~s$^{-1}$ for the $3\,Z_\odot$ case. Expressed relative to the bolometric luminosity, these values are $L_\mathrm{kin}/L_\mathrm{bol} \simeq 7.8$--$12\%$ and $\simeq 2.3$--$3.6\%$, respectively. For an inflow, this quantity represents the kinetic energy flux carried inward with the gas rather than the feedback power commonly associated with UFOs.

For comparison, the accretion rate required to power the observed bolometric luminosity at a standard radiative efficiency $\eta=0.1$ \citep[e.g.,][]{marconi04,hopkins07,davis_laor11} is $\dot{M}_\mathrm{acc}=L_\mathrm{bol}/(\eta c^2)\simeq 1.5\,M_\odot$~yr$^{-1}$. At solar metallicity, $\dot{M}_\mathrm{in}/\dot{M}_\mathrm{acc} \simeq 0.6$--1.0, indicating that the inflow may carry enough mass to sustain the observed AGN luminosity, while at $3\,Z_\odot$ we obtain $\dot{M}_\mathrm{in}/\dot{M}_\mathrm{acc} \simeq 0.2$--0.3, still a significant fraction of the accretion budget. In either case, under the adopted assumptions, the local inward mass flux could therefore make an important contribution to AGN feeding, although an inflow rate inferred at these radii does not establish that all of the material ultimately reaches the SMBH.

\subsection{Cascade scenario and physical interpretation}
\label{sec:disc_cascade}

The repeated detection of the primary UFI over the May~2023, August~2024, and January~2026 epochs must be interpreted against the much shorter dynamical timescales of the gas. For the shell-like geometry adopted above, the characteristic crossing time is $t_\mathrm{cross}=R/v_\mathrm{in}$. At the inferred characteristic radius, $R\simeq49$--$77\,R_\mathrm{g}$, this gives $t_\mathrm{cross}\simeq3$--$5$~days. The free-fall time from rest, including the reduction of the inward acceleration by radiation pressure, is
\begin{equation} 
t_\mathrm{ff}=\frac{\pi}{2\sqrt{2}}\left(\frac{R^3}{G M_\mathrm{BH} P_\mathrm{rad}}\right)^{1/2},
\end{equation}
with the pure-gravity limit recovered for $P_\mathrm{rad}=1$. This ranges from $t_\mathrm{ff}\simeq4.9$~days at $R\simeq49\,R_\mathrm{g}$ when radiation pressure is included ($P_\mathrm{rad}=0.64$), to $t_\mathrm{ff}\simeq7.8$~days at $R\simeq77\,R_\mathrm{g}$ in the pure free-fall case. In what follows, we refer to both these timescales together as the dynamical timescale of the inflow.
Over the 2.7~yr baseline of the three \nustar\ detections (2.2~yr in the source rest frame), the absorbing material would therefore have crossed this region, or been replenished, hundreds of times. The absorber is thus unlikely to be a single long-lived cloud. Instead, our results point to a continuously replenished inflow in which new material of similar velocity and ionization is resupplied on the few-day dynamical timescale, possibly feeding the AGN in an accretion ``cascade''.

The thermal state of the absorber provides an additional physical constraint on the nature of the inflow. As detailed in Appendix~\ref{app:stability}, we computed photoionization stability curves, which map the equilibrium temperature of the gas against pressure ionization parameter, for different assumptions about the ionizing continuum. In no case is the absorber securely thermally stable, being at best marginally stable or unstable. At the characteristic radius, the conservative estimate described in the same Appendix gives a thermal-adjustment time of $t_\mathrm{th}\simeq20$--$50$~min, hundreds of times shorter than the dynamical timescale. The gas therefore has ample time to respond to marginal or unstable conditions within a single infall, making a structured or multiphase flow physically plausible. This thermal susceptibility is compatible with, but does not by itself establish, the continuously replenished inflow inferred from the dynamical timescales.

The 2010 \suzaku\ non-detection provides an earlier constraint on the current inflow episode. Dedicated MC simulations (Appendix~\ref{app:suzaku}) show that an absorber with properties similar to those inferred in this work would have produced detectable absorption in the \suzaku\ spectrum. However, no absorption is observed at the predicted $\sim4.8$~keV energy, ruling out such an absorber at $>3\sigma$ under the adopted assumptions. We note that this test does not exclude an inflow with a harder ionizing continuum, a different ionization state, a lower column density, or different kinematics, any of which could weaken the Fe~K absorption.
The UFI absorption signature was therefore either absent during the \suzaku\ epoch or too weak to be detected. It has since recurred across a source-rest-frame baseline of at least 2.2~yr. If it was absent in 2010, it emerged within the subsequent 10.3~yr in the source rest frame, corresponding to an elapsed interval of at most 12.5~yr between its onset and the 2026 observation.

With these dynamical, thermal, and temporal constraints in mind, we consider several physical scenarios that could account for the observed inflow. In the Chaotic Cold Accretion framework \citep[CCA;][]{gaspari13,gaspari17,gaspari20}, turbulence and radiative cooling of the hot gaseous halo drive the condensation of warm and cold clouds. These clouds lose angular momentum through cloud--cloud and cloud--torus collisions \citep[e.g.,][]{pizzolato_soker10} and rain stochastically toward the nucleus \citep[e.g.,][]{barbani26a,cammelli26a,piana26a}. Applied to ESP~39607, the UFI could represent the innermost observable stage of this process, with the recurrence arising from a sequence of clumps or filaments continuously supplied by the larger-scale accretion flow. This interpretation naturally explains the contrast between the multi-year recurrence and the day-scale dynamical time. Depending on the assumed ionizing continuum, the absorber is consistent with either marginal stability or active thermal instability. Its rapid thermal response relative to the infall further supports a clumpy, continuously replenished interpretation, with the Fe\,\textsc{xxv} He$\alpha$/Fe\,\textsc{xxvi} Ly$\alpha$ absorption possibly tracing a highly ionized envelope or transition layer associated with denser clumps supplied by a broader CCA feeding cascade.

Another possibility is a magnetically guided, non-equatorial relativistic inflow. \citet{fukumura26} proposed such a model partly motivated by the UFI first reported in ESP~39607 by \citetalias{peca25a}. Building on earlier non-equatorial accretion-flow calculations \citep[e.g.,][]{fukumura07}, the model identifies UFIs with weakly magnetized, hydro-dominated accretion flows following non-equatorial streamlines above the disc, organized by a large-scale poloidal magnetic field. The calculation is performed in general relativistic hydrodynamics in Kerr spacetime and predicts redshifted Fe~K absorption from gas reaching $v\gtrsim0.1c$ within $\sim100\,R_\mathrm{g}$, comparable to the radii and velocities inferred here. \citet{fukumura26} further suggest that, if the inflow originates at larger disc radii, it could produce a persistent absorption signature over multi-year timescales. In this interpretation, the consistency of the line energy across the three epochs may trace a persistent inflow channel, while the short dynamical time requires the absorbing gas within that channel to be continuously replaced. The thermal susceptibility inferred here remains compatible with this picture, since such a channel does not require individual gas clumps to remain thermally stable. Indeed, the stability curves provide a first-order description, while magnetic pressure and anisotropic conduction could further regulate the formation and survival of such structures.

Additional interpretations include misaligned or disrupted disc configurations, failed disc winds, and aborted jets. Disc tearing can produce discrete rings or fragments that lose angular momentum and plunge inward on eccentric or counter-rotating trajectories \citep[e.g.,][]{nixon12,dogan18}, qualitatively consistent with multiple inflowing components. In a failed-wind scenario, gas lifted from the disc becomes overionized, loses radiative acceleration before escaping, and falls back toward the disc \citep[e.g.,][]{proga04}. This could naturally connect the tentative inflowing and outflowing components as different stages of the same wind. In the aborted-jet picture, intermittently ejected blobs with sub-escape velocities reach a maximum radius and subsequently fall back toward the SMBH \citep[e.g.,][]{ghisellini04}, potentially allowing inflow and outflow signatures to coexist. However, these scenarios are more naturally associated with episodic fallback, making the recurrence of the primary UFI at a consistent energy across all three epochs less straightforward to explain.

The dynamical, thermal, and temporal constraints therefore favor a continuously replenished inner inflow, but they do not uniquely determine its physical origin. The CCA and magnetically guided interpretations need not be mutually exclusive. A broader CCA cascade could supply clumpy material from larger radii, while a persistent field-guided channel could organize its inward motion through the inner $\sim100\,R_\mathrm{g}$. In such a scale-linked picture, the UFI would trace the innermost portion of the AGN feeding chain. However, as noted in Section~\ref{sec:disc_energetics}, we stress that a mass flux measured at these radii does not establish that all the material is ultimately accreted. 
At the same time, fallback from a failed wind or aborted jet remains a viable alternative, particularly in view of the candidate UFO component. 
More broadly, the secondary UFI and UFO candidates, together with the observed flux decline and moderately high Eddington ratio, suggest that the primary inflow may be embedded in a complex environment where inflow and outflow coexist \citep[e.g.,][]{king15,giustini19}. 

Deeper \xrism\ observations will be needed to test the tentative components and determine whether the primary absorber contains multiple ionization phases at the same velocity. Continued X-ray monitoring will constrain the duration and duty cycle of the inflow, test for changes in its velocity, and measure its ionization response to continuum variability.
In addition, millimeter and optical/infrared observations could test whether this inner inflow is connected to a larger-scale multiphase feeding flow.

\section{Summary and Conclusions} \label{sec:conclusions}

We presented the first high-resolution X-ray spectroscopic observation of the ultra-fast inflow candidate in ESP~39607 with \xrism, complemented by simultaneous \xmm\ and \nustar\ data. Our main findings are:

\begin{itemize}

    \item The high spectral resolution of \resolve\ separates the previously blended \nustar\ feature into two absorption components near 4.7 and 4.9~keV in the observed frame. A single \textsc{xstar} absorber identifies them with Fe\,\textsc{xxv} He$\alpha$ and Fe\,\textsc{xxvi} Ly$\alpha$ at $z_\mathrm{abs}\simeq0.413$, corresponding to an inflow velocity of $v_\mathrm{in}\simeq0.16c$. The preferred solution uses the $v_\mathrm{turb}=1000$~km~s$^{-1}$ grid and gives $\log\xi/\mathrm{erg~s^{-1}~cm}\simeq3.7$--$3.8$ and $\log N_\mathrm{H,abs}/\mathrm{cm}^{-2}\simeq23.2$--$23.8$, where the range in column density is driven by the assumed metallicity. These measurements support and refine the UFI interpretation proposed by \citetalias{peca25a} from the earlier \nustar\ data.

    \item The primary-UFI detection is robust against the choice of baseline continuum model. Across \textsc{MYTorus}, \textsc{X-skirtor}, and \textsc{uxclumpy}, the linked Gaussian pair gives Gaussian-equivalent fit improvements of $3.1$--$3.5\sigma$ and $\Delta\mathrm{AIC}\simeq9$--12, while the physically motivated \textsc{xstar} models give $2.9$--$3.4\sigma$ and $\Delta\mathrm{AIC}\simeq7$--11. Using MC simulations, the primary pair reaches a significance of $3.7\sigma$. Combining this result with the independent 2023--2024 MC evidence gives an overall significance of $5.3\sigma$ against the joint no-feature hypothesis.

    \item The absorber velocity, ionization, column density, line energy, and equivalent width are consistent with the earlier \nustar\ measurements, indicating recurrent UFI absorption across three epochs. A simple chance-alignment calculation gives a probability of only a few percent that three unrelated features would occur within the observed energy range. 
    At the inferred characteristic radius of $R\simeq49$--$77\,R_\mathrm{g}$, the dynamical timescale is only $\sim3$--8~days. Moreover, under all tested assumptions about the ionizing continuum, the absorber is never securely thermally stable, being at best marginally stable or unstable, with a thermal-adjustment time of only $t_\mathrm{th}\simeq20$--$50$~min. Together, the multi-epoch recurrence and the short dynamical and thermal timescales favor a continuously replenished inner inflow over a single long-lived cloud. The \suzaku\ non-detection indicates that an absorption signature like those observed in 2023, 2024, and 2026 was either absent or too weak to detect in 2010. If genuinely absent, the feature emerged between December~2010 and its first detection in May~2023, constraining the interval between its onset and the 2026 observation to 2.2--12.5~yr in the source rest frame.

    \item Under the assumed geometry, the inferred mass inflow rate is $\dot{M}_\mathrm{in}\simeq0.9$--$1.4\,M_\odot$~yr$^{-1}$ at solar abundance and $\simeq0.27$--$0.42\,M_\odot$~yr$^{-1}$ at $3\,Z_\odot$. These values correspond to $\dot{M}_\mathrm{in}/\dot{M}_\mathrm{acc}\simeq0.6$--$1.0$ and $\simeq 0.2$--0.3, respectively. The UFI may therefore make an important contribution to AGN feeding, potentially as the innermost stage of an accretion cascade.

\end{itemize}

Beyond the primary UFI, we find tentative evidence for two weaker absorption components, a secondary UFI at $v_\mathrm{in}\sim0.31c$ ($2.8\sigma$) and a UFO at $v_\mathrm{out}\sim0.08$--$0.12c$ ($2.0\sigma$). Together, these tentative inflowing and outflowing signatures suggest that the primary UFI may be embedded in a dynamically complex environment. Additionally, the neutral Fe~K$\alpha$ profile may comprise a narrow core ($\lesssim1000$~km~s$^{-1}$) and a broader component with an FWHM of $\sim7600$~km~s$^{-1}$, consistent with an origin in the broad-line region or inner torus wall.

These results make ESP~39607 a particularly valuable target for high-resolution X-ray spectroscopy. It hosts one of the clearest UFI candidates known and, to our knowledge, the only such candidate identified in a Type~2 AGN, providing a rare view of highly ionized gas moving inward at tens of gravitational radii. Its recurrence at a consistent characteristic energy across three epochs, despite dynamical timescales of only days, favors sustained replenishment over a single long-lived structure or isolated fallback events. Despite this evidence, whether the inflow originates in a chaotic cold accretion cascade, a magnetically guided channel, or the fallback of material that fails to escape the gravitational potential remains an open question. Distinguishing among these scenarios will require continued monitoring and deeper high-resolution spectroscopy, as discussed in Section~\ref{sec:disc_cascade}.

\begin{acknowledgments}
We acknowledge the anonymous referee for the valuable comments that improved the quality of the paper.
AP thanks the \xmm\ and \nustar\ teams for the coordination with the \xrism\ observation. AP thanks D. Costanzo for the useful discussions. AP and MK acknowledge support from the NASA grants 80NSSC26K0159 and 80NSSC26K0901.
AP acknowledges support from the Agencia Nacional de Investigación y Desarrollo (ANID) through ANID/FONDECYT/Postdoctorado 3260848.
RS acknowledges funding from the CAS-ANID grant number CAS220016.
MG acknowledges support from the ERC Consolidator Grant \textit{BlackHoleWeather} (101086804).
KF is grateful to the College of Science and Mathematics at JMU for financial support.
GC acknowledges that part of this work was performed at JPL/Caltech under NASA contract 80NM0018D0004.
\end{acknowledgments}





\facilities{\xrism, \xmm, \nustar, \suzaku}

\software{
          HEASoft,
          SAS \citep{SAS},
          XSPEC \citep{xspec},
          XSTAR \citep{xstar1,xstar2},
          Matplotlib \citep{matplotlib}.
          }

\appendix

\section{Line characterization and multi-epoch variability} \label{app:linechar_var}
\hypertarget{appendix.A}{\mbox{}}

\subsection{Best-fit line parameters from the phenomenological 2-Gaussian model}

We report in Table~\ref{tab:lines} the best-fit observed-frame energies and equivalent widths of the primary Fe\,\textsc{xxv} He$\alpha$/Fe\,\textsc{xxvi} Ly$\alpha$ pair, obtained with the two-Gaussian model and free centroid energies (see Section~\ref{sec:gaussians} for details), for each of the three baseline continuum models. The line energies are mutually consistent within $\sim$10~eV across the three baseline continuum models, and the equivalent widths are consistent within their uncertainties.

\begin{deluxetable*}{lcccc}
\tabletypesize{\footnotesize}
\tablecaption{Best-fit observed-frame energies and equivalent widths of the primary absorption pair from the two-Gaussian model with $\sigma=10$~eV and free centroid energies. \label{tab:lines}}
\tablewidth{0pt}
\tablehead{
\colhead{Model} & \colhead{$E_\mathrm{Fe\,XXV}$} & \colhead{$\mathrm{EW}_\mathrm{Fe\,XXV}$} & \colhead{$E_\mathrm{Fe\,XXVI}$} & \colhead{$\mathrm{EW}_\mathrm{Fe\,XXVI}$} \\
\colhead{} & \colhead{(keV)} & \colhead{(eV)} & \colhead{(keV)} & \colhead{(eV)}
}
\startdata
\textsc{MYTorus}   & $4.67 \pm 0.01$ & $53^{+19}_{-13}$   & $4.92 \pm 0.01$ & $30^{+26}_{-4}$ \\
\textsc{X-skirtor}  & $4.66 \pm 0.02$ & $44^{+18}_{-4}$  & $4.92 \pm 0.01$ & $34^{+14}_{-7}$ \\
\textsc{uxclumpy} & $4.67^{+0.02}_{-0.01}$ & $51^{+10}_{-16}$   & $4.92 \pm 0.01$ & $34^{+21}_{-7}$  \\
\enddata
\end{deluxetable*}

\subsection{NuSTAR consistency across three epochs}

To compare the absorption feature consistently across the three epochs, we re-analyzed the January 2026 \nustar-only data using the same procedure as \citetalias{peca25a} for the May 2023 and August 2024 observations. The spectra were grouped to a minimum of 5 counts per bin and fitted with the same \textsc{uxclumpy}-based continuum model. We then added a single unresolved Gaussian absorption line ($\sigma=10$~eV) with free centroid energy. The resulting line energies and equivalent widths of the blended $\sim$4.8~keV feature are listed in Table~\ref{tab:nustar_epochs} together with the earlier epochs.
For reference, the joint 2026 fit with \xrism, \xmm, and \nustar\ resolves this feature into two components, with values reported in Table~\ref{tab:lines}, whereas the \nustar-only fit measures their blended centroid. Given its broad uncertainty, the \nustar-only centroid is broadly consistent with the energy range spanned by the two resolved components. As the \nustar-only equivalent width represents the combined absorption of the blended pair, we do not compare it with the equivalent widths of the two individually resolved components.

\begin{figure}[ht]
    \centering
    \includegraphics[width=\columnwidth]{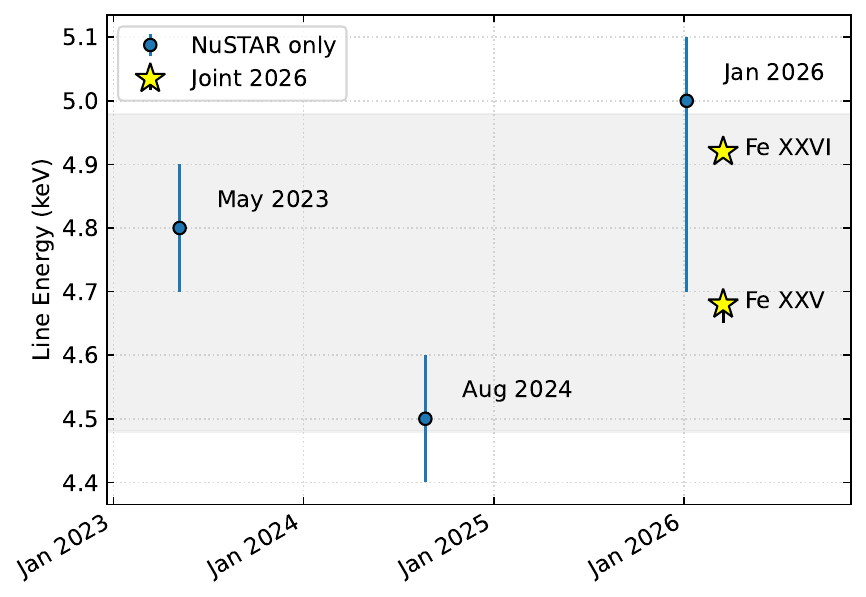}
    \caption{Observed-frame absorption-line energies across the three \nustar\ epochs. Blue circles show the \nustar-only single-Gaussian measurements, with the 2023 and 2024 values from \citetalias{peca25a}. Yellow stars show the two components resolved in the joint 2026 fit. The gray band shows the centroid uncertainties from the MC simulations of \citetalias{peca25a}.}
    \label{fig:line_changes}
\end{figure}

\begin{figure}[ht]
    \centering
    \includegraphics[width=\columnwidth]{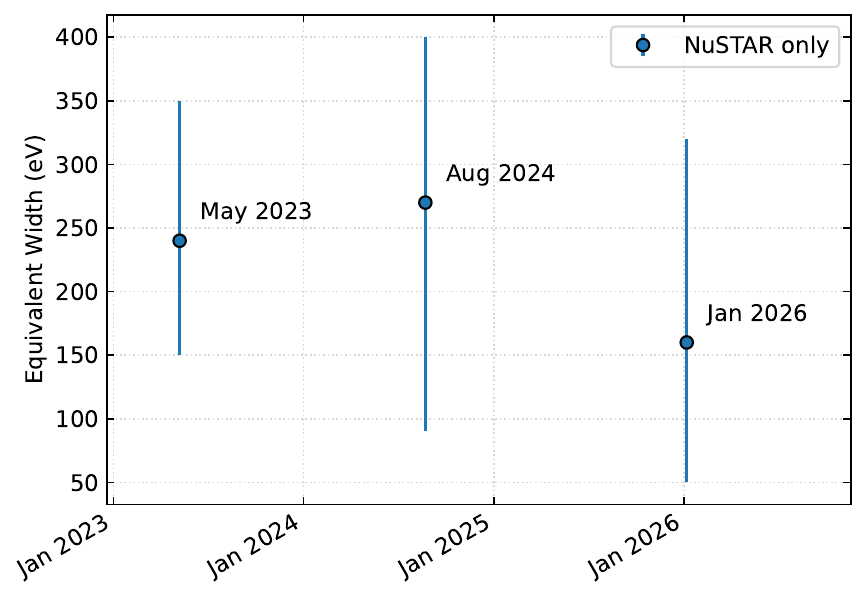}
    \caption{Equivalent widths of the absorption feature across the three \nustar\ epochs of ESP~39607. The \nustar-only equivalent widths are consistent within uncertainties across all epochs, indicating no significant variation.}
    \label{fig:eqw_changes}
\end{figure}

Figures~\ref{fig:line_changes} and~\ref{fig:eqw_changes} show the evolution of the \nustar-only centroid and equivalent width, with Figure~\ref{fig:line_changes} also showing the two components resolved in the joint 2026 fit. \citetalias{peca25a} found the 2023 and 2024 centroid measurements to be consistent at the 99\% confidence level. The gray band shows the centroid range derived by \citetalias{peca25a} from Monte Carlo simulations of a 2023-like feature observed at the lower flux and signal-to-noise ratio of the 2024 observation. Accounting for these Monte Carlo uncertainties, the line centroids are consistent across all three epochs, as are the equivalent widths. We therefore find no significant evidence for variation in either quantity.

\begin{deluxetable}{lccc}
\tabletypesize{\footnotesize}
\tablecaption{\nustar-only absorption-line parameters across the three epochs, obtained with an unresolved Gaussian line over the baseline \textsc{uxclumpy} continuum. Net exposures are the averages from the FPMA/FPMB cameras. \label{tab:nustar_epochs}}
\tablewidth{0pt}
\tablehead{
\colhead{Epoch} & \colhead{Net exposure} & \colhead{$E_\mathrm{line}$} & \colhead{EW} \\
\colhead{} & \colhead{(ks)} & \colhead{(keV)} & \colhead{(eV)}
}
\startdata
2023 May & 21.1 & $4.8 \pm 0.1$ & $240^{+110}_{-90}$ \\
2024 Aug & 21.5 & $4.5 \pm 0.1$ & $270^{+130}_{-180}$ \\
2026 Jan & 27.8 & $5.0^{+0.1}_{-0.3}$ & $160^{+160}_{-110}$ \\
\enddata
\end{deluxetable}

\section{Extended model-comparison grid} \label{app:fullgrid}

We report in Table~\ref{tab:modelcomp_full} the full grid of phenomenological and physically motivated model variants cited in the main text. 
As in Table~\ref{tab:modelcomp}, $\Delta\mathcal{C}$/$\Delta$dof are computed relative to the corresponding baseline continuum model, except for the lower block, which is relative to the best-fit, single-zone \textsc{xstar} baseline with solar abundances and $v_\mathrm{turb}=1000$~km~s$^{-1}$.

\begin{deluxetable*}{lccccccccc}
\tabletypesize{\footnotesize}
\tablecaption{Extended model grid described in Section~\ref{sec:spectral}, complementing Table~\ref{tab:modelcomp}. The primary-UFI blocks test Gaussian variants, including linked-energy pairs with $\Delta E_\mathrm{rest}=0.266$~keV, and \textsc{xstar} grids with the indicated turbulent velocities. The secondary-UFI blocks similarly test double-Gaussian and \textsc{xstar} variants, while the UFO block uses a Gaussian with $\sigma=10$~eV. These models are compared with their corresponding baseline continua. The lower block tests additional kinematic components relative to the best-fit solar-abundance, single-zone \textsc{xstar} model with $v_\mathrm{turb}=1000$~km~s$^{-1}$. AIC labels follow Table~\ref{tab:modelcomp}, extended here to include modest support (M) for $0<\Delta\mathrm{AIC}<4$ and no support (N) for $\Delta\mathrm{AIC}\leq0$. \label{tab:modelcomp_full}}
\tablewidth{0pt}
\tablehead{
\colhead{Model variant} &
\multicolumn{3}{c}{\textsc{MYTorus}} &
\multicolumn{3}{c}{\textsc{X-skirtor}} &
\multicolumn{3}{c}{\textsc{uxclumpy}} \\
\colhead{} &
\colhead{$\Delta\mathcal{C}$/$\Delta$dof} & \colhead{Sig.} & \colhead{$\Delta$AIC} &
\colhead{$\Delta\mathcal{C}$/$\Delta$dof} & \colhead{Sig.} & \colhead{$\Delta$AIC} &
\colhead{$\Delta\mathcal{C}$/$\Delta$dof} & \colhead{Sig.} & \colhead{$\Delta$AIC}
}
\startdata
\multicolumn{10}{c}{\textit{Primary UFI: phenomenological Gaussian variants}}\\
$+$\,1 broad Gaussian             & $14.0/3$ & $3.0\sigma$ & $8.0$ / S & $11.8/3$ & $2.6\sigma$ & $5.8$ / P  & $14.5/3$ & $3.1\sigma$ & $8.5$ / S \\
$+$\,2 Gaussians, fixed $\sigma$, free $E$   & $22.0/4$ & $3.7\sigma$ & $14.0$ / VS & $17.7/4$ & $3.2\sigma$ & $9.7$ / S  & $21.8/4$ & $3.7\sigma$ & $13.8$ / VS \\
$+$\,2 Gaussians, fixed $\sigma$, linked $E$ & $16.4/3$ & $3.3\sigma$ & $10.4$ / VS & $14.7/3$ & $3.1\sigma$ & $8.7$ / S  & $17.5/3$ & $3.5\sigma$ & $11.5$ / VS \\
$+$\,2 Gaussians, free $\sigma$, free $E$    & $22.1/6$ & $3.2\sigma$ & $10.1$ / VS & $19.0/6$ & $2.9\sigma$ & $7.0$ / S  & $21.8/6$ & $3.2\sigma$ & $9.8$ / S \\
$+$\,2 Gaussians, free $\sigma$, linked $E$  & $20.1/5$ & $3.2\sigma$ & $10.1$ / VS & $16.2/5$ & $2.7\sigma$ & $6.2$ / P  & $20.3/5$ & $3.3\sigma$ & $10.3$ / VS \\
\hline
\multicolumn{10}{c}{\textit{Primary UFI: \textsc{xstar} turbulent-velocity variants}}\\
$+$\,\textsc{xstar}, $v_\mathrm{turb}=100$~km~s$^{-1}$  & $14.6/3$ & $3.1\sigma$ & $8.6$ / S   & $8.2/3$ & $2.0\sigma$ & $2.2$ / M  & $15.5/3$ & $3.2\sigma$ & $9.5$ / S \\
$+$\,\textsc{xstar}, $v_\mathrm{turb}=1000$~km~s$^{-1}$ & $15.7/3$ & $3.2\sigma$ & $9.7$ / S   & $13.3/3$ & $2.9\sigma$ & $7.3$ / S  & $17.1/3$ & $3.4\sigma$ & $11.1$ / VS \\
$+$\,\textsc{xstar}, $v_\mathrm{turb}=5000$~km~s$^{-1}$ & $12.6/3$ & $2.8\sigma$ & $6.6$ / P   & $11.0/3$ & $2.5\sigma$ & $5.0$ / P  & $13.5/3$ & $2.9\sigma$ & $7.5$ / S  \\
\hline
\multicolumn{10}{c}{\textit{Secondary UFI: phenomenological Gaussian variants}}\\
$+$\,2 Gaussians, fixed $\sigma$, free $E$   & $16.3/4$ & $3.0\sigma$ & $8.3$ / S   & $16.7/4$ & $3.1\sigma$ & $8.7$ / S  & $14.1/4$ & $2.7\sigma$ & $6.1$ / P   \\
$+$\,2 Gaussians, fixed $\sigma$, linked $E$ & $10.5/3$ & $2.4\sigma$ & $4.5$ / P   & $10.0/3$ & $2.4\sigma$ & $4.0$ / P      & $10.2/3$ & $2.4\sigma$ & $4.2$ / P   \\
$+$\,2 Gaussians, free $\sigma$, free $E$    & $17.7/6$ & $2.7\sigma$ & $5.7$ / P   & $17.1/6$ & $2.6\sigma$ & $5.1$ / P  & $17.7/6$ & $2.7\sigma$ & $5.7$ / P   \\
$+$\,2 Gaussians, free $\sigma$, linked $E$  & $14.3/5$ & $2.5\sigma$ & $4.3$ / P   & $14.1/5$ & $2.4\sigma$ & $4.1$ / P  & $14.5/5$ & $2.5\sigma$ & $4.5$ / P   \\
\hline
\multicolumn{10}{c}{\textit{Secondary UFI: \textsc{xstar} turbulent-velocity variants}}\\
$+$\,\textsc{xstar}, $v_\mathrm{turb}=100$~km~s$^{-1}$  & $9.8/3$ & $2.3\sigma$ & $3.8$ / M & $7.3/3$ & $1.9\sigma$ & $1.3$ / M & $10.8/3$ & $2.5\sigma$ & $4.8$ / P \\
$+$\,\textsc{xstar}, $v_\mathrm{turb}=1000$~km~s$^{-1}$ & $15.3/3$ & $3.2\sigma$ & $9.3$ / S & $12.3/3$ & $2.7\sigma$ & $6.3$ / P & $16.2/3$ & $3.3\sigma$ & $10.2$ / VS \\
$+$\,\textsc{xstar}, $v_\mathrm{turb}=5000$~km~s$^{-1}$ & $14.6/3$ & $3.1\sigma$ & $8.6$ / S & $11.4/3$ & $2.6\sigma$ & $5.4$ / P & $15.2/3$ & $3.2\sigma$ & $9.2$ / S \\
\hline
\multicolumn{10}{c}{\textit{UFO: phenomenological Gaussian}}\\
$+$\,1 Gaussian, fixed $\sigma$              & $5.5/2$ & $1.9\sigma$ & $1.5$ / M & $6.3/2$ & $2.0\sigma$ & $2.3$ / M & $7.1/2$ & $2.2\sigma$ & $3.1$ / M \\
\hline
\multicolumn{10}{c}{\textit{Additional fits relative to the single-zone \textsc{xstar} baseline}}\\
$+$\,\textsc{xstar} phase, same $v_\mathrm{in}$ and $v_\mathrm{turb}$     & $3.9/2$ & $1.5\sigma$ & $-0.1$ / N & $3.2/2$ & $1.3\sigma$ & $-0.8$ / N & $3.8/2$ & $1.4\sigma$ & $-0.2$ / N \\
$+$\,\textsc{xstar} secondary UFI, 1000~km~s$^{-1}$              & $10.1/3$ & $2.4\sigma$ & $4.1$ / P & $7.7/3$ & $1.9\sigma$ & $1.7$ / M & $9.9/3$ & $2.3\sigma$ & $3.9$ / M \\
$+$\,UFO Gaussian, fixed $\sigma$                              & $4.5/2$ & $1.6\sigma$ & $0.5$ / M & $5.2/2$ & $1.8\sigma$ & $1.2$ / M & $6.0/2$ & $2.0\sigma$ & $2.0$ / M       \\
\enddata
\end{deluxetable*}

\section{\resolve-only analysis of the primary UFI} \label{app:resolve}

\begin{figure}[ht]
    \centering
    \includegraphics[width=\columnwidth]{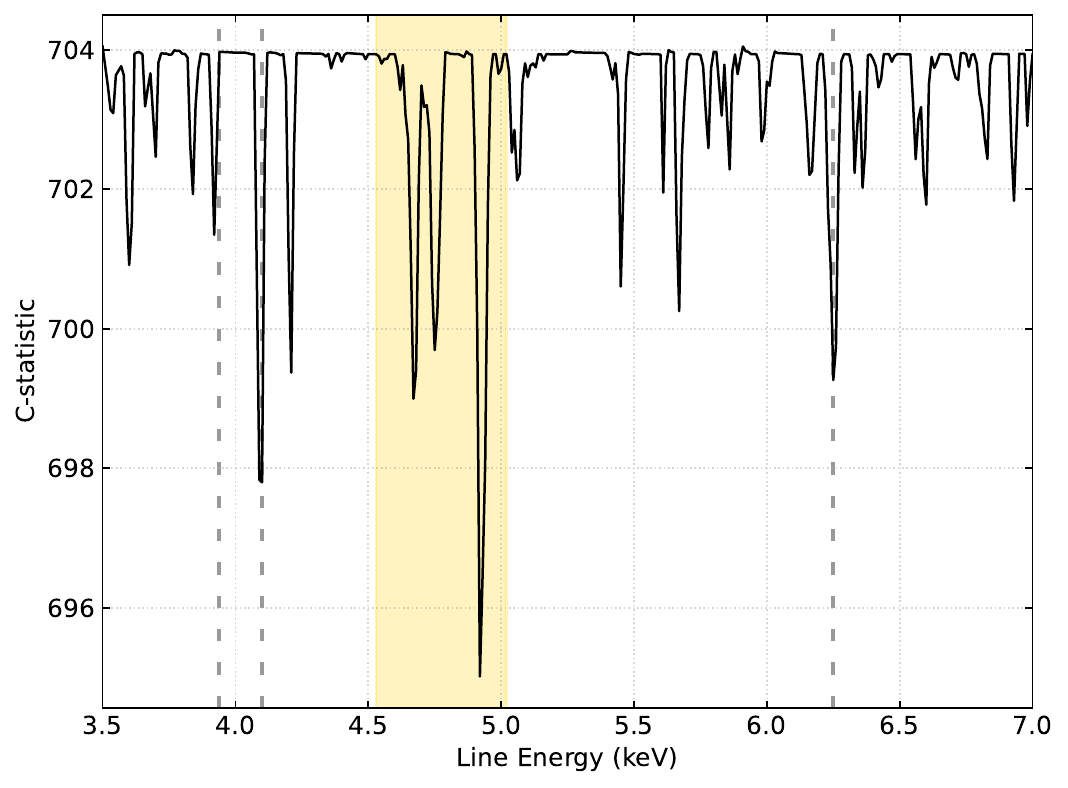}
    \caption{One-dimensional blind scan over the centroid energy of a Gaussian absorption line ($\sigma=10$~eV), performed on the \resolve\ spectrum alone with the \textsc{MYTorus} model. The gold shaded band marks the energy interval containing the primary UFI pair, while the gray dashed lines indicate the energies of the secondary UFI and UFO candidates, as in Figure~\ref{fig:noabs}.
    All energies are shown in the observed frame.}
    \label{fig:resolve_scan}
\end{figure}

We repeated the analysis of Sections~\ref{sec:gaussians} and~\ref{sec:xstar} using only the \resolve spectrum. The three baseline models were retained, with the same free continuum parameters except the scattering fraction, which was fixed to its joint-fit value (Table~\ref{tab:xstar}) because the \resolve fit is restricted to energies above 3.5~keV, where the secondary power-law contribution is negligible. 
We performed a blind one-dimensional scan over the centroid energy of a Gaussian absorption line, using the \textsc{MYTorus} baseline continuum and allowing the line normalization to vary \citep[e.g.,][]{peca21,reeves26}. This recovers the primary UFI as the deepest minimum in the band (Figure~\ref{fig:resolve_scan}). We also find weaker minima at the energies of the candidate secondary UFI and UFO identified in Section~\ref{sec:secondary}. Although the scan is noisier, as expected from the reduced photon statistics, the \resolve spectrum alone is sensitive to the same absorption structures seen in the joint fit.

As in Section~\ref{sec:gaussians}, we then fitted the fiducial linked Gaussian pair, with $\sigma=10$~eV and a fixed rest-frame separation of $\Delta E_\mathrm{rest}=0.266$~keV. Relative to the corresponding \resolve-only baseline continua, the pair improves the fits by $\Delta\mathcal{C}/\Delta$dof $=10.3$--$11.0/3$ (2.4--2.5$\sigma$), with $\Delta\mathrm{AIC}=4.3$--5.0, indicating ``positive'' evidence for all three continuum models. 
Leaving both centroid energies free to vary gives $\Delta\mathcal{C}/\Delta$dof $=10.7$--$13.3/4$, corresponding to 2.2--2.6$\sigma$, with $\Delta\mathrm{AIC}=2.7$--5.3. The best-fit energies, $\simeq4.67\pm0.02$ and $\simeq4.92\pm0.01$~keV, and the corresponding equivalent widths, of $\simeq30$--45~eV and $\simeq40$--60~eV, are consistent with the joint-fit values of Table~\ref{tab:lines} within their uncertainties. The \resolve data alone therefore recover both components at the energies measured in the joint analysis.

We then fitted the fiducial solar-abundance \textsc{xstar} grid ($v_\mathrm{turb}=1000$~km~s$^{-1}$), which gives $\Delta\mathcal{C}/\Delta$dof $=10.0$--$14.4/3$ (2.4--3.0$\sigma$) and $\Delta\mathrm{AIC}=4.0$--8.4, indicating ``positive'' to ``strong'' evidence for the two-line interpretation. This yields $z_\mathrm{abs}=0.412\pm0.004$, $\log N_\mathrm{H,abs}/\mathrm{cm}^{-2}\simeq23.8$--24.1, and $\log\xi/\mathrm{erg~s^{-1}~cm}\simeq3.6$--3.8, consistent with the joint-fit solution (Table~\ref{tab:xstar}) and hence $v_\mathrm{in}\simeq0.16c$. The continuum parameters are likewise consistent with the joint-fit values in Table~\ref{tab:xstar}, though with uncertainties a factor of $\sim$3--6 larger.

We also attempted to describe the pair with a single broad Gaussian, as done in Section~\ref{sec:gaussians}, leaving the line width free to vary. No stable broad solution is found for any of the three continua. For all models, the fits converge onto one of the two narrow components. This is expected, as in the joint fit the CCD spectra cannot resolve the two components, which therefore appear as a single blended feature, leaving a broad-line description formally viable (Table~\ref{tab:modelcomp_full}). \resolve instead resolves the pair and does not admit a single broad solution. Finally, we repeated the MC simulations of Section~\ref{sec:gaussians} using only the \resolve response and exposure, with $10^4$ realizations, obtaining a significance of 2.1$\sigma$. 

The lower significances presented in this section, compared to those obtained from the joint fit, are expected, since \resolve contributes only $\sim15\%$ of the counts used in the joint analysis and the broadband data are needed to constrain the obscured continuum. Nevertheless, the \resolve-only result independently supports the two-line interpretation.

\section{Thermal stability of the inflow}
\label{app:stability}

The location of photoionized gas on a thermal-equilibrium stability curve depends on the shape of the ionizing continuum. We therefore tested whether the thermal stability of the primary UFI is sensitive to the photon index adopted in the \textsc{xstar} calculation. In addition to the fiducial $\Gamma=2$ grid, we recomputed the solar-abundance grid using $\Gamma=1.7$, representative of the average best-fitting continuum slope (Table~\ref{tab:xstar}), and repeated the fits for each continuum model.
For the three baseline models, the fit statistic increases by only $\Delta\mathcal{C}=0.3$--1.4 relative to the fiducial grid. The fits are therefore statistically indistinguishable, and the primary-UFI detection is preserved in all three continuum models.
With the harder grid, the absorber redshift and inflow velocity remain unchanged. The \textsc{MYTorus}, \textsc{X-skirtor}, and \textsc{uxclumpy} fits all give $\log\xi=3.0^{+0.6}_{-0.3}$ and $\log N_\mathrm{H,abs}/\mathrm{cm}^{-2}=23.4^{+0.5}_{-0.3}$. As expected, the harder continuum favors a lower ionization parameter and a corresponding shift in column density, although the confidence intervals overlap those obtained with the fiducial grid. 
Thus, the detection and inferred kinematics of the primary UFI are not significantly affected by the photon index assumed for the ionizing continuum.

We then computed thermal-equilibrium curves with \textsc{xstar} for four intrinsic ionizing continua. The first two are power laws with $\Gamma=2$ and $\Gamma=1.7$, matching the continua used to generate the fiducial and harder \textsc{xstar} grids, respectively. To explore the effects of a more realistic broadband spectral energy distribution, we also generated two \textsc{agnsed} continua \citep{kubota18} with the same slopes. Unlike a simple power law, \textsc{agnsed} includes optical--UV disc emission and warm Comptonization, providing a more complete description of the radiation that sets the heating and cooling balance. We adopted $\log(M_\mathrm{BH}/M_\odot)=8.26$, $L_\mathrm{bol}/L_\mathrm{Edd}=0.36$, and $i=70\degr$, following Sections~\ref{sec:continuum} and \ref{sec:disc_energetics}.

Thermal-equilibrium stability curves are conventionally expressed as equilibrium temperature versus $\xi/T$, where $\xi/T$ is proportional to the pressure ionization parameter. Positive-slope branches are locally stable to isobaric perturbations, negative-slope branches are unstable, and turning points indicate marginal stability \citep{krolik81,chakravorty09}. Such curves have also been used to assess the thermal stability of highly ionized AGN outflows, including UFOs \citep[e.g.,][]{kraemer18,xu25}. In those studies, UFOs are typically found on stable branches, which can support persistence and a high duty cycle \citep[e.g.,][]{reeves26}.

\begin{figure}[ht]
    \centering
    \includegraphics[width=\columnwidth]
    {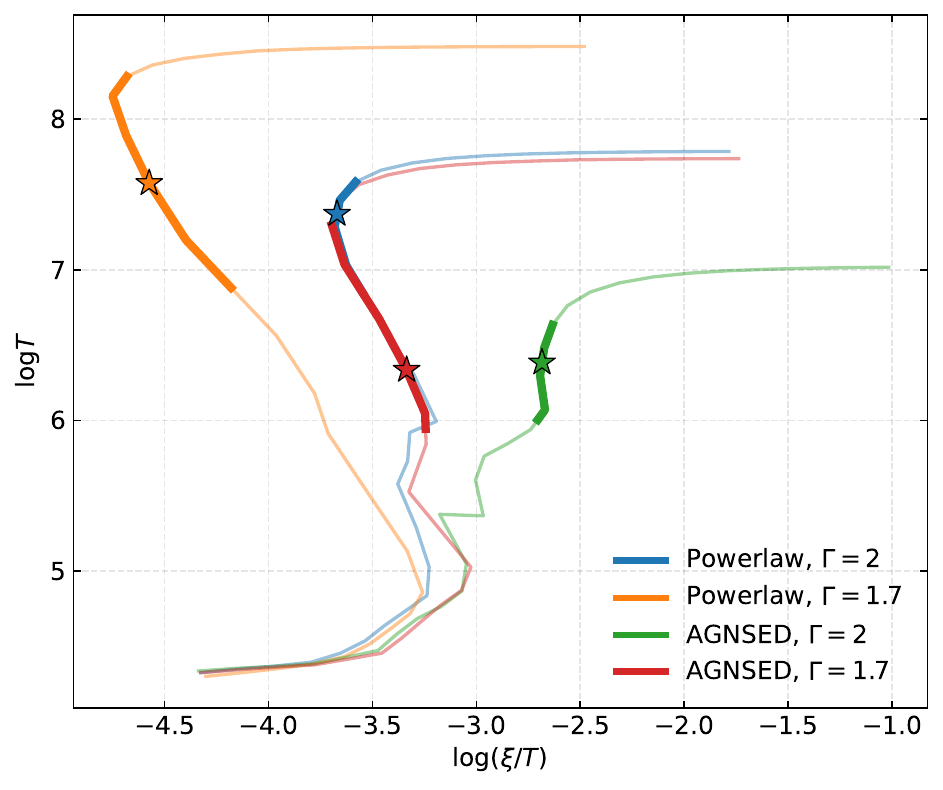}
    \caption{Thermal-equilibrium stability curves for the primary UFI. Blue and orange curves show the assumed power-law continua with $\Gamma=2$ and $\Gamma=1.7$, respectively, while green and red curves show the corresponding \textsc{agnsed} continua. Stars mark the absorber locations corresponding to the representative ionization parameters, $\log\xi=3.7$ and $\log\xi=3.0$, for $\Gamma=2$ and $\Gamma=1.7$, respectively. Thick segments indicate the ranges spanned by the corresponding ionization-parameter uncertainties across the continuum models. In no case is the absorber securely located on a positive-slope branch.}
    \label{fig:scurve}
\end{figure}

Figure~\ref{fig:scurve} shows the four curves obtained for the power-law and \textsc{agnsed} continua. To locate the absorber on each curve, we used representative ionization parameters averaged over the continuum models, $\log\xi=3.7$ for $\Gamma=2$ and $\log\xi=3.0$ for $\Gamma=1.7$. For both continuum prescriptions, the $\Gamma=2$ solution lies at a turning point, whereas the $\Gamma=1.7$ solution lies on a negative-slope branch. The absorber is therefore either marginally stable or thermally unstable, depending on the assumed photon index, and is never securely located on a positive-slope branch.
As an additional check, we repeated the stability-curve calculation using \textsc{kynsed} continua \citep{dovciak22} with the same two photon indices. The resulting curves are similar to the corresponding \textsc{agnsed} curves and yield the same stability classification, and are therefore not shown in Figure~\ref{fig:scurve} for visual clarity.

To estimate a characteristic thermal-adjustment time without assuming a gas density, we use the Compton energy-exchange timescale appropriate to the upper, Compton-dominated branch of the stability curves  \citep[e.g.,][]{sazonov04}:
\begin{equation}
\label{eq:tth}
t_\mathrm{th}\simeq\frac{3\pi}{2}\,
\frac{m_\mathrm{e}c^{2}R^{2}}
{\sigma_\mathrm{T}L_\mathrm{bol}},
\end{equation}
where $m_\mathrm{e}$ is the electron mass and $\sigma_\mathrm{T}$ is the Thomson cross-section. At $R\simeq49$--$77\,R_\mathrm{g}$, this gives $t_\mathrm{th}\simeq20$--$50$~min, hundreds of times shorter than the dynamical timescale of $\sim3$--8~days derived in Section~\ref{sec:disc_cascade}. At the lower temperatures obtained for the tested continua, additional atomic heating and cooling may shorten the response time. We therefore regard the Compton timescale as a conservative upper estimate of the characteristic thermal-adjustment time.

The gas can therefore adjust its thermal state many times during a single infall, making its location on the stability curve dynamically relevant. Indeed, combined with the absence of a clearly thermally stable solution, this supports a picture in which the inflow can develop an inhomogeneous or clumpy structure rather than remain a single smooth, long-lived component. When considered together with the multi-epoch recurrence at a consistent velocity, these properties favor continuously supplied scenarios, such as a CCA accretion cascade or a magnetically guided inflow, rather than episodic fallback of material initially launched outward, as discussed in Section~\ref{sec:disc_cascade}.

\section{\suzaku\ non-detection of the primary UFI} \label{app:suzaku}

While \citetalias{peca25a} found no evidence for absorption in the December 2010 \suzaku\ spectrum, here we use MC simulations to test whether the absorption signature of a UFI with the properties inferred in 2026 would have been detectable in those data. We used the \suzaku XIS0, XIS1, and XIS3 spectra extracted by \citet{ricci17}, grouped to a minimum of 20 counts per bin, and fitted them with the same \textsc{uxclumpy} continuum adopted throughout this work and in \citetalias{peca25a}.

We generated two sets of $2000$ \texttt{fakeit} realizations from the best-fit \suzaku\ continuum model. The first set was generated from the continuum alone, providing the null distribution expected in the absence of the UFI. In the second set, we additionally imposed the single-zone \textsc{xstar} absorber inferred from the 2026 joint fit, thereby testing \suzaku's sensitivity to a UFI with the properties inferred in 2026. For these absorber-injected simulations, we adopted the best-fit solar-abundance parameters, $N_\mathrm{H,abs}\simeq6\times10^{23}$~cm$^{-2}$, $\log\xi\simeq3.7$, $z_\mathrm{abs}=0.413$, and $v_\mathrm{turb}=1000$~km~s$^{-1}$ (Table~\ref{tab:xstar}). Because the absorber redshift is fixed by the 2026 solution, each realization in both sets was then fitted with and without a redshifted Fe\,\textsc{xxv} He$\alpha$/Fe\,\textsc{xxvi} Ly$\alpha$ absorption pair at the observed-frame energies predicted by the best-fit absorber redshift. We fixed $\sigma=10$~eV and allowed only the two line normalizations to vary.

We then fitted the real \suzaku\ spectrum with and without the absorption pair fixed at the energies predicted by the 2026 solution. The addition of the two lines produced no improvement in the fit ($\Delta\mathcal{C}\simeq0$), with both line normalizations consistent with zero, indicating no evidence for absorption at the predicted UFI energies. We compared this result with the two simulated data sets. None of the $2000$ absorber-injected realizations yielded a $\Delta\mathcal{C}$ as small as that measured in the real spectrum, implying that a UFI with the column density and ionization inferred in 2026 is disfavored at $>3\sigma$, with the significance limited by the finite number of simulations. Additionally, the absorption pair was recovered at $>2\sigma$ in $75.0\%$ of the absorber-injected realizations, demonstrating that \suzaku\ had substantial sensitivity to a 2026-like UFI. Conversely, $96.9\%$ of the continuum-only realizations yielded a line improvement as small as or smaller than that observed in the real spectrum, confirming that the negligible $\Delta\mathcal{C}$ measured in the data is entirely consistent with a spectrum containing no UFI absorption. We nevertheless note that this constraint is conditional on the adopted absorber model, as an inflow with a lower column density or different ionization state could have remained undetected. We therefore conclude that, under these assumptions, the 2010 \suzaku\ observation strongly disfavors the presence of a UFI with the properties inferred in 2026.

\bibliography{sample701}{}
\bibliographystyle{aasjournalv7}

\end{document}